\documentclass[fleqn,usenatbib,useAMS]{mnras}

\usepackage[usestackEOL]{stackengine}
\usepackage{multirow}

\usepackage{graphicx}	%
\usepackage{amsmath}	%
\usepackage{amssymb}	%
\usepackage{multicol}        %
\usepackage{bm}		%
\usepackage{pdflscape}	%

\usepackage[T1]{fontenc}
\usepackage{ae,aecompl}

\usepackage{newtxtext,newtxmath}

\title[WL mass--$\mu_{\star}$ calibration for clusters]{
Weak-lensing calibration of a stellar mass-based mass proxy for redMaPPer and Voronoi Tessellation clusters in SDSS Stripe 82}

\author[M. E. S. Pereira et al.]{Maria E. S. Pereira$^{1,2}$\thanks{Contact e-mail: \href{mailto:mariaeli@brandeis.edu}{mariaeli@brandeis.edu}},
Marcelle Soares-Santos$^{1,3}$,
Martin Makler$^{2}$,
James Annis$^{3}$,
\newauthor Huan Lin$^{3}$, 
Antonella Palmese$^{4,3}$,
Andr\'e Z. Vitorelli$^{5}$, 
Brian Welch$^{6,3}$,
\newauthor Gabriel B. Caminha$^{7,8}$,
Thomas Erben$^{9}$, 
Bruno Moraes$^{4}$ and 
Huanyuan Shan$^{9}$ 
\\
$^{1}$Brandeis University, 415 South Street, Waltham, MA 02453, USA\\
$^{2}$Coordena\c{c}\~ao de Cosmologia, Astrof\'{\i}sica e Intera\c{c}\~oes Fundamentais, Centro Brasileiro de Pesquisas F\'{\i}sicas, Rio de Janeiro, RJ 22290-180, Brasil\\
$^{3}$Center for Particle Astrophysics, Fermi National Accelerator Laboratory, Batavia, IL 60510, USA\\
$^{4}$Department of Physics and Astronomy, University College London, Gower Street, London, WC1E 6BT, UK\\
$^{5}$Instituto de Astronomia, Geof\'isica e Ci\^encias Atmosf\'ericas, Departamento de Astronomia, S\~ao Paulo-SP, 05508-090,  Brasil\\
$^{6}$Department of Physics, The University of Chicago, Chicago, IL 60637, USA \\
$^{7}$Dipartimento di Fisica e Scienze della Terra, Universit\`a degli Studi di Ferrara, Via Saragat 1, I-44122 Ferrara, Italy \\
$^{8}$INAF - Osservatorio Astronomico di Bologna, Via Ranzani 1, I- 40127
Bologna, Italy \\
$^{9}$Argelander-Institut f\"ur Astronomie, Auf dem H\"ugel 71, 53121 Bonn, Germany
}

\date{}

\pubyear{2017}

\begin{document}
\label{firstpage}
\pagerange{\pageref{firstpage}--\pageref{lastpage}}
\maketitle

\begin{abstract} %
We present the first weak lensing  calibration of $\mu_{\star}$, a new galaxy cluster mass proxy corresponding to the total stellar mass of red and blue members, in two cluster samples selected from the SDSS Stripe 82 data: 230 redMaPPer clusters at redshift $0.1\leq z<0.33$ and 136 Voronoi Tessellation (VT) clusters at $0.1 \leq z < 0.6$. We use the CS82 shear catalogue and stack the clusters in $\mu_{\star}$ bins to measure a mass-observable power law relation. For redMaPPer clusters we obtain $M_0 = (1.77 \pm 0.36) \times 10^{14}h^{-1} M_{\odot}$, $\alpha = 1.74 \pm 0.62$. For VT clusters, we find $M_0 = (4.31 \pm 0.89) \times 10^{14}h^{-1} M_{\odot}$, $\alpha = 0.59 \pm 0.54$ and $M_0 = (3.67 \pm 0.56) \times 10^{14}h^{-1} M_{\odot}$, $\alpha = 0.68 \pm 0.49$ for a low and a high redshift bin, respectively. Our results are consistent,  internally and with the literature, indicating that our method can be applied to any cluster-finding algorithm. In particular, we recommend that $\mu_{\star}$ be used as the mass proxy for VT clusters. Catalogs including $\mu_{\star}$ measurements will enable its use in
studies of galaxy evolution in clusters and cluster cosmology.
\end{abstract}


\newpage

\section{Introduction}

Galaxy clusters are the largest and most massive gravitationally bound structures in the Universe. They are formed by a large number of galaxies (usually with one large elliptical central), hot gas and dark matter evolving in strongly coupled processes. Cluster properties depend on both the dynamical processes that take place inside them and on the evolution of the Universe. As such, they can be used as a powerful tool to probe its content, to study the formation and evolution of structures, and to test modified gravity theories \citep{2001ApJ...553..545H,RevModPhys.77.207,2011ARA&A..49..409A,2012ARA&A..50..353K,2013SSRv..177....1E,2014JCAP...05..039P,2015Sci...347.1462H,2016ApJ...825L...1M,2016JCAP...04..023P}.

Galaxy clusters also act as powerful gravitational lenses. Their intense gravitational fields produce distortions in the shape (shear) of the background galaxies (sources). Through this effect, we can assess the mass distribution of the galaxy clusters to use them as cosmological tools \citep{2005astro.ph..9252S}. At the depths of ongoing and planned wide-field surveys, it is not possible to measure this signal from individual clusters, except for the most massive ones. However, we can combine the lensing signal of a large number of clusters to obtain a higher signal-to-noise. This {\it stacking} procedure requires the large statistics enabled by wide-field surveys such as the Dark Energy Survey\footnote{\url{https://www.darkenergysurvey.org/}} (DES; \citealt{2016MNRAS.460.2245J, 2016arXiv161006890M}), the Canada--France--Hawaii--Telescope (CFHT) Lensing Survey\footnote{\url{http://www.cfhtlens.org/}} (CFHTLens; \citealt{2014MNRAS.437.2111V, 2015MNRAS.447.1304F, 2015MNRAS.451.1460K}), the Sloan Digital Sky Survey (SDSS; {\citealt{2001ApJ...554..881S, 2012ApJ...748..128S, 2015MNRAS.452..701W, 2017MNRAS.465.1348G, 2017MNRAS.466.3103S}}), and the Kilo-Degree Survey\footnote{\url{http://kids.strw.leidenuniv.nl/}} (KiDS; \citealt{2013Msngr.154...44J, 2015MNRAS.454.3500K}). 

Clusters can be identified in several wavelengths such as in X-rays, radio and optical. In particular, the identification in the optical can be made through the search for overdensities (from matched-filters to more complex Voronoi tessellations) of multi-band optically detected galaxies. These multi-band optical cluster catalogues usually provide good cluster photometric redshifts (photo-z), which are crucial information for weak lensing measurements. 

Observationally, galaxy clusters are ranked not by the mass of the halo but by some proxy for mass. A mass-observable relation must be calibrated in order to make the connection between the observable and the true halo mass. The technique of stacking the weak lensing signal for many systems within a given observable interval provides one of the most direct and model independent ways to accurately calibrate such mass-observable scaling relations. Many efforts have been made to determine the scaling relations empirically using an observable mass proxy for the cluster mass. However, comparing the empirical measurements is challenging since there are several methods to identify the clusters, which lead to different cluster samples, and different definitions of the mass proxy to be used \citep{2007arXiv0709.1159J, 2014MNRAS.444..147O, 2015MNRAS.447.1304F, 2015ApJ...807..178W, 2015MNRAS.452..701W, 2017MNRAS.466.3103S}. 

In this work, we use the stacked weak lensing technique on galaxy clusters identified by two different algorithms to estimate their mass and to obtain the scaling relations for two different mass proxies. The clusters are identified by the red-sequence Matched-filter Probabilistic Percolation\footnote{\url{https://github.com/erykoff/redmapper}}  \citep[redMaPPer;][]{2014ApJ...785..104R} optical cluster finder and the geometric Voronoi Tessellation\footnote{\url{https://github.com/soares-santos/vt}} algorithm \citep[VT;][]{2011ApJ...727...45S} in the Sloan Digital Sky Survey (SDSS) Stripe 82 region. We use the weak lensing shear catalog from the CFHT Stripe 82 Survey \citep[CS82;][]{2014RMxAC..44..202M,erben2017}, which has excellent image quality and thus we expect our mass estimates to be less affected by shape systematics than the results obtained from the SDSS data alone (see, e.g. Gonzalez et al. in prep.). In our analysis, we obtain the scaling relations for both the redMaPPer optical richness $\lambda$ \citep{2012ApJ...746..178R,2014ApJ...785..104R} and for a new mass proxy $\mu_{\star}$, which is described in two companion papers \citep{welch2017,palmese2017}.

The new mass proxy $\mu_{\star}$ is defined as the sum of the stellar masses of cluster galaxies weighted by their membership probabilities. This quantity can be estimated reliably from optical photometric surveys \citep{2016MNRAS.463.1486P} and shows a tight correlation with the total cluster mass \citep[e.g.][]{2012A&A...548A..83A}. \cite{palmese2017} perform a matching between redMaPPer DES clusters and XMM X-ray clusters at $0.1<z<0.7$ and demonstrate that $\mu_{\star}$ has low scatter with respect to X-ray mass observables. They compute the $T_X$--$\mu_{\star}$ relation, obtaining a scatter of $\sigma_{\ln T_X|\mu_{\star}} = 0.20$, which is comparable with results found for the redMaPPer richness estimator $\lambda$ by \citet{2016ApJS..224....1R} using XMM and Chandra X-ray samples at $0.2<z<0.9$ and by \citet{2014ApJ...783...80R} using the XCS X-ray sample at $0.1<z<0.5$. 

When using the redMaPPer mass-proxy $\lambda$, we obtain a $M_{200}$--$\lambda$ relation that is consistent with previous measurements found in the literature. When using $\mu_{\star}$ on the same sample our results show a similar level of uncertainty. Our results for the VT sample in the same redshift range are consistent (within $1.5 \sigma$) with those we obtain with redMaPPer, showing that our mass calibration is robust against the specifics of the cluster selection algorithms. Finally, we extend our analysis to a higher redshift VT sample. We do not see an evolution of the mass-observable relation at the level of precision of this analysis.

This paper is organized as follows. In Section \ref{data}, we describe the cluster and the lensing shear catalogues. In section \ref{methodology} we present the methodology for the measurement and modelling of the stacked cluster masses. We present our results and the derived mass-calibrations in Section \ref{results}. Finally, in Section \ref{conclusions} we present our concluding remarks. In this paper, the distances are expressed in physical coordinates, magnitudes are in the AB system (unless otherwise noted) and we assume a flat $\Lambda$CDM cosmology with $\Omega_{m}=0.3$ and $H_0 = 100\,h\,\mathrm{km\,s^{-1} Mpc^{-1}}$.

\section{Catalogs in SDSS Stripe 82} 
\label{data}

We work with data on the so-called Stripe 82 region, which is an equatorial stripe that has been scanned multiple times as part of the SDSS supernovae search \citep{2008AJ....135..338F}, leading to a  5-band co-add of selected images about two magnitudes deeper than the main SDSS survey \citep{2014ApJ...794..120A}. Stripe 82 has become a well studied $\sim$100 sq-deg scale region, with extensive spectroscopy from SDSS and other wide-field spectroscopic surveys \citep{6dF,wigglez,2SLAQ,2QZ,2dFGRS,2004MNRAS.349.1397C,2009MNRAS.392...19C,2011AJ....142...72E}, reaching fainter magnitudes in smaller regions \citep{Garilli08,DEEP2,PRIMUS,VIPERS,2013A&A...559A..14L}, and a large spectral coverage from several synergistic surveys \citep[see, e.g. ][and references therein]{S82X21, SpIES1, 2017arXiv170505451G}, including NIR photometry from UKIRT Infrared Deep Survey \citep[UKIDSS,][]{UKIDSS} and from a combination of CFHT WIRCam and Visible and Infrared Survey Telescope for Astronomy (VISTA) VIRCAM data \citep{2017arXiv170505451G}. It serves as a precursor of future datasets, and is being covered by ongoing surveys at higher depths (e.g. DES, HSC\footnote{\url{http://hsc.mtk.nao.ac.jp/ssp/}}) and denser wavelength coverage (J-PLUS\footnote{\url{https://confluence.astro.ufsc.br:8443/}}; Mendes de Oliveira et al. in prep.).

In this work we use three catalogues in Stripe 82: 
\begin{enumerate}

\item A cluster catalog resulting from the VT algorithm \citep{2011ApJ...727...45S} applied to the SDSS  
stripe 82 co-add catalog \citep{2014ApJ...794..120A}
with neural network photometric redshift measurements \citep{2012ApJ...747...59R}.
\item A catalog of galaxy clusters identified by the redMaPPer algorithm \citep{2014ApJ...785..104R} on the SDSS  \textit{8th Data Release} \citep[DR8;][]{2011ApJS..193...29A}.
\item A galaxy catalog from the CS82 survey \citep{erben2017} including shape measurements and photometric redshifts from  matched SDSS co-add \citep{2014ApJ...794..120A} and UKIDSS YJHK \citep{UKIDSS} photometry. 
\end{enumerate}

The CS82 survey defines the sky footprint of our analysis and both cluster catalogues are matched to it. The CS82 photo-zs were computed by \citet{2015ApJS..221...15B} with the BPZ code \citep{2000ApJ...536..571B}. These are more precise than previously available photo-z measurements \citep[see also][]{2017MNRAS.467.3024L, Soo17} and therefore we use them throughout this analysis, namely, 
in the computation of membership probabilities, for determining absolute magnitudes, and in the stacked weak lensing analysis. 

\subsection{VT clusters} 
\label{vt}
The VT cluster finder  \citep{2011ApJ...727...45S}
uses a geometric technique to construct Voronoi cells that contain only one object each. The cell sizes are inversely proportional to the local density and a galaxy cluster candidate is defined as a high-density region composed of small adjacent cells. The raw number of member galaxies, $N_{\mathrm{VT}}$, is thus the number of VT cells. The key point is to estimate the density threshold to separate an overdensity (a galaxy cluster) from the background and take into account the projection effects due to the fact that the Voronoi cells are computed in a 2D-distribution of objects in the sky. In order to achieve that, the VT algorithm is built in photo-z shells and uses the two-point correlation function of the galaxies in the field to determine the density threshold for detection of the cluster candidates and their significance. Since it is a geometric technique, there is no need of \textit{a priori} assumption on galaxy colours, the presence of a red-sequence or any assumptions about their astrophysical properties.

In this paper, we use the VT catalogue produced for the Stripe 82 co-add \cite[v1.10;][]{2015MNRAS.452..701W}. Since that release version, the VT team has developed an improved membership assignment scheme and a new mass proxy, $\mu_{\star}$. In this work we incorporate those developments (see section \ref{mu_star_section} for details) and add two new improvements, namely, a defragmentation algorithm and a redefinition of the cluster central galaxy (described in sections \ref{defrag_section} and  \ref{cen_section}, respectively). The former mitigates the effect of photometric redshift shell edges and of multiple density peaks within individual clusters. The latter
allows us to extend the probabilistic approach of membership to the determination of the central cluster galaxy.

\subsubsection{Assigning the new mass proxy $\mu_{\star}$}
\label{mu_star_section}
$N_{\mathrm{VT}}$ performs poorly as a mass proxy, as shown by the scatter in the richness-mass relation presented  in \cite{2015MNRAS.454.2305S}. The new mass proxy, $\mu_{\star}$, is based on a probabilistic membership assignment scheme \citep{welch2017}\footnote{\url{https://github.com/bwelch94/Memb-assign}} and on measurements of stellar masses \citep{palmese2017}\footnote{\url{https://github.com/apalmese/BMAStellarMasses}}. 
In particular, \citet{palmese2017} showed that the scatter in the $\mu_\star$ to X-ray temperature relation is comparable to that of other mass proxies for an X-ray selected sample and that it allows interesting cluster evolution analyses, having a clear physics meaning of the cluster stellar mass. 

The first step in computing $\mu_{\star}$ is to compute the membership probability $P_{\mathrm{mem}}$ for each cluster galaxy
\begin{equation}
P_{\mathrm{mem}}=P_z \cdot P_{r} \cdot P_{c},
\end{equation}
where the three components represent the probability of the galaxy being a member given its redshift ($P_z$), its distance from the cluster centre ($P_r$) and its colour ($P_{c}$):
\begin{itemize}
\item $P_z$ is the integrated redshift probability distribution of each galaxy within a $\Delta$$z=0.1$ window of the cluster. 
\item $P_r$ is computed assuming a projected Navarro-Frenk-and-White profile. 
\item $P_c$ is determined via Gaussian 
mixture modeling of the galaxy colour distribution with two components, red sequence and blue cloud; it is defined as the sum of the probability that the galaxy colour is drawn from either the blue or red component.
\end{itemize} 

For membership assignment purposes we use a subsample of the galaxy catalog {cut } at $M_r < -19$. That subsample is volume limited over our redshift range.
We calculated the absolute magnitudes using kcorrect v4\_2 \citep{2007AJ....133..734B} taking the BPZ photo-z as the galaxy redshift. We constructed a grid of $g-r$, $r-i$, and $i-z$ colours from the templates in kcorrect and chose the closest to the observed galaxy colours. That chosen template provides the K-correction 
from observed $i$ band to rest-frame $r$-band, which, together with our chosen cosmology, allows us to calculate $M_r$. 

After computing the membership probabilities for each galaxy $i$ within 3 Mpc of each cluster $j$, we compute their stellar masses assuming that every member galaxy is at the redshift of its host, $M_{\star,i}(z_j)$. Because the cluster redshifts have smaller uncertainties than individual galaxies, this minimizes the uncertainties on $M_{\star,i}$ measurements. Stellar masses are computed using a Bayesian model averaging method (BMA, see e.g. \citealt{hoeting}). With this method, we take into account the uncertainty on model selection by fitting a set of robust, up to date stellar population synthesis (SPS) models and averaging over all of them. In this work we use the flexible stellar population synthesis (FSPS) code by \citet{fsps} to generate simple stellar population spectra. Those are computed assuming Padova (\citealt{padova1}, \citealt{padova2}, \citealt{padova3}) isochrones  and Miles (\citealt{miles}) stellar libraries with four different metallicities ($Z=0.03,0.09,0.0096$ and 0.0031). We choose the four-parameter star formation history described in \citet{simha}. Finally, once the stellar masses are computed, we define the new mass proxy as the sum of the individual galaxy stellar masses weighted by their membership probability:   
\begin{equation}
\mu_{\star} = \sum_{i} P_{\mathrm{mem},i} M_{\star,i} \,.%
\end{equation}

The membership assignment and $\mu_{\star}$ computation methods were applied only to VT clusters with $N_{\mathrm{VT}}>20$, to avoid poorly detected galaxy groups. 
After applying the CS82 mask and a photometric redshift cut at $z<0.6$, where the VT sample is most reliable, we obtain a sample of 136 clusters, which are used throughout this analysis. 

\subsubsection{Investigating cluster fragmentation} \label{defrag_section}
Fragmentation of large clusters into smaller components in the VT catalogue is one of the sources of scattering in the observable-mass relation. We uncovered the issue by performing cylindrical matching (angular separation $\theta < 1$ arcmin and $\Delta z < 0.05$) between redMaPPer and VT catalogs. This comparison showed some cases where one redMaPPer cluster was split into two or more VT clusters.
 
When applied to a cluster fragment, the new probabilistic membership method will result in a full-fledged list of members, as the probabilities are computed out to 3 Mpc radius. This is a designed feature. For two fragments located near each other, the result will be two instances of the same cluster with slightly different membership probabilities. In that case, only one instance should be maintained in the catalogue. In order to ensure that, we developed a defragmentation method using the membership probabilities $P_{\mathrm{mem}}$. For a given pair of cluster candidates, we define the "true" cluster as the one for which $\sum P_{\mathrm{mem}}$ is the largest. 

In practice we first attribute a flag for each cluster in the catalog as if they were all unique real clusters (\texttt{cluster\_frag}=1). Then, we rank them by mass proxy and compute the angular separation between each other. If the separation is smaller than the largest $R_{200}$ between the two and the redshift difference is $\Delta z<0.05$, those clusters are considered to be two instances, $i$ and $j$, of a fragmented pair. We compute the summation of the member probabilities of the fragmented clusters $i$ and $j$ as $P_i = \sum P_{\mathrm{mem}}^i$ and $P_j = \sum P_{\mathrm{mem}}^j$, respectively. We then match their members list (in our membership scheme, clusters may share members) and then compute the quantity $P_{\mathrm{match}} = \sum P_{\mathrm{mem}}^{i,\mathrm{match}}=\sum P_{\mathrm{mem}}^{j,\mathrm{match}}$ for the matched members. Once we have these quantities we compute the fractions   
\begin{equation}
f_{ij} = \frac{P_{\mathrm{match}}}{P_i}\,\,\, \mathrm{and}\,\,\,f_{ji} = \frac{P_{\mathrm{match}}}{P_j}.
\end{equation}
Since $P_{\mathrm{match}}$ is the same for both, the only difference is in the denominator. If $f_{ij} \leq f_{ji}$, then $i$ is kept in the catalog while $j$ is 
removed (i.e. set \texttt{cluster\_frag}=0). %
We apply this procedure to VT clusters in the range $0.1 \leq z <0.6$ and we find that $\sim 16$ per cent of the clusters were affected by this issue. This is therefore a non-negligible correction and future versions of VT catalog should have this new procedure applied to them before being released.  

\subsubsection{Redefining the cluster central galaxy} \label{cen_section}
The brightest cluster galaxy (BCG) is a good proxy for the 
centre of the cluster and that fact is used in several cluster finding methods \citep[e.g.,][]{2007ApJ...660..239K, 2010ApJS..191..254H, 2014MNRAS.444..147O}. The original VT algorithm, however, takes a purely spatial approach and defines the cluster central galaxy as the one inside the highest density VT cell. After computing $\mu_{\star}$ we redefine the central cluster galaxy as the member galaxy with 
maximum probability of membership. 
The probability  $P_{\mathrm{cen}}^{\star}$ 
that this newly defined central galaxy 
is the true centre of the cluster is proportional to 
its membership probability:
\begin{equation}
P_{\mathrm{cen}}^{\star} \propto \mathrm{max}(P_{\mathrm{mem}}).
\end{equation}
Although not normalized, this centring probability is analogous to that of the redMaPPer algorithm.

\subsection{redMaPPer clusters}
\label{redmapper}

The redMaPPer cluster finder \citep{2014ApJ...785..104R} uses multi-band colours to find overdensities of red-sequence galaxies around candidate central galaxies. In SDSS data, redMaPPer uses the five band magnitudes ($ugriz$) and their errors to spatially group the red-sequence galaxies at similar redshifts into cluster candidates. For each red galaxy, redMaPPer estimates its membership probability ($p_{\mathrm{mem}}$) following a matched-filter technique. At the end, for each identified cluster, redMaPPer will return an optical richness estimate $\lambda$ (the total sum of the $p_{\mathrm{mem}}$ of all galaxies that belong to that cluster), a photo-z estimate $z_{\lambda}$, and the positions and probabilities of the five most likely central galaxies ($P_{\mathrm{cen}}$).

In this work we use the most recent version of the SDSS redMaPPer public catalog \cite[v6.3; ][]{2016ApJS..224....1R}, which covers an area of $10^4 \,\, \mathrm{deg^2}$, down to a limiting magnitude of $i=21$ for galaxies. The full sample of redMaPPer clusters in the catalog has $0.08 \lesssim z_{\lambda} <0.6$ and $ 20 \lesssim \lambda < 300$. 
After restricting the catalog to the $\sim 170 \,\, \mathrm{deg^2}$ of the CS82 footprint, we restrict our mass measurements to the low redshift bin $0.1\leq z_{\lambda} <0.33$ to enable comparison with previous SDSS weak lensing measurements and because the redMaPPer cluster catalog from single epoch SDSS data is most reliable at these redshifts. The redMaPPer sample used in this work, after all selection criteria are applied, contains 230 clusters.

We compute $\mu_{\star}$ as well for the redMaPPer clusters,
employing the same steps described in section \ref{mu_star_section}.
This means that new membership probabilities are computed for every cluster and enables direct comparison between the $\Delta\Sigma$ profiles obtained for $\lambda$ and $\mu_{\star}$, as discussed in section \ref{rm-results}. The defragmentation step was not needed for redMaPPer.

\subsection{CS82 weak lensing catalog}
\label{cs82}

We use the shape measurements from the CS82 survey, which is a joint Canada--France--Brazil project using MegaCam at CFHT and is specially designed to study the weak and strong lensing effects \citep{erben2017}. The survey has 173 MegaCam pointings in the $i^{\prime}$ band covering an effective area of $127\,\,\mathrm{deg^2}$ (after masking to avoid bright stars, satellite tracks and other image artefacts) to a limiting magnitude of $24$ and mean seeing of $0.6$ arcsec \citep{2017MNRAS.467.3024L} providing excellent imagining quality for precise shape measurements. The shape estimates were obtained with Lensfit code \citep{2007MNRAS.382..315M} that performs a Bayesian profile-fitting of the surface brightness to obtain an unbiased estimate of the shear components from the average ellipticities. The code was tested in simulations and real data \citep{2008MNRAS.390..149K, 2013MNRAS.429.2858M}, achieving very good results \citep{2012MNRAS.423.3163K} and became a suitable tool for precise shape estimates in surveys with the imagining quality of CS82. 

Lensfit was applied to the masked imaging data following the same pipeline as the CFTHLenS collaboration \citep{2013MNRAS.429.2858M} and applying the shear calibration factors and testing the systematics in the same way as \cite{2012MNRAS.427..146H}. For each source, an additive calibration correction factor $c_2$ is applied to the $\epsilon_2$ shear component and a multiplicative shear calibration factor as a function of the signal-to-noise ratio and size of the source, $m(\nu_{\mathrm{SN}},r)$, is also computed. Besides that, the Lensfit shear measurements were also compared with other independent shear calibration methods \citep{2012MNRAS.425.2610R, 2014MNRAS.440.2922M, 2015MNRAS.454.3357C} by \cite{2017MNRAS.467.3024L} who have found that a largely unknown and unaccounted for bias in the Lensfit measurements is an unlikely possibility. From the Lensfit output catalog we select the objects with weight $w>0$, $\mathrm{FITCLASS}=0$ and $\mathrm{MASK} \leq 1$. These quantities are computed by Lensfit, where $w$ is an inverse variance weight for each source, $\mathrm{FITCLASS}$ is a star/galaxy separation flag to remove stars and select galaxies with well-measured shapes and $\mathrm{MASK}$ is a flag that indicates the quality of the photometry, where for most of the weak lensing analysis $\mathrm{MASK} \leq 1$ is a robust cut to apply, as shown by \citet{2013MNRAS.433.2545E}. We also select only galaxies with magnitudes $20 \leq i^{\prime} \leq 24.7$, with the upper value corresponding to the limit to which the shear measurements were accurately calibrated in the CFHT images \citep{2012MNRAS.427..146H, 2013MNRAS.429.2858M}.  

The BPZ photometric redshift catalogue includes, in addition to the photo-zs and errors, the parameter $\mathrm{BPZ\_ODDS}$ that varies between 0 to 1 and indicates catastrophic redshift errors. We removed from our source galaxy sample all objects with $\mathrm{BPZ\_ODDS}\leq 0.5$. According to \cite{2012MNRAS.421.2355H} and \cite{2013MNRAS.431.1547B} the photo-z of the sources degrade at $z_s > 1.3$, which could be a concern for our measurements. However, \cite{2017MNRAS.467.3024L} performed a test computing the CS82 lensing signal with and without this redshift cut and have shown no statistically significant shift in the signal. Therefore we do not apply any restriction on the maximum value of $z_s$ so as to maximize the number of background sources. Finally, after applying all the aforementioned cuts we obtained a final catalogue with 2~809~764 sources, which give an effective weighted galaxy number density of $n_{\mathrm{eff}} = 4.5$ galaxies $\mathrm{arcmin^{-2}}$.          

Previous weak lensing measurements using the CS82 source catalog have been performed, e.g. by \citet{2014MNRAS.442.2534S,2014MNRAS.438.2864L,2015PhRvD..91f2001H,2015MNRAS.450.2888L, 2016MNRAS.458.2573L,2016JCAP...08..013B, 2017MNRAS.467.3024L,2017ApJ...840..104S, 2017arXiv170303348N}, making this lensing catalog well tested for different applications.

\section{Methodology}
\label{methodology}

We measure the mass-observable relation from the stacked lensing signal of redMaPPer and VT clusters using the CS82 shear catalogue. For the stacking of the lenses, we define bins of redshift and observable mass proxy.
 
\begin{figure}
 \includegraphics[width=\columnwidth]{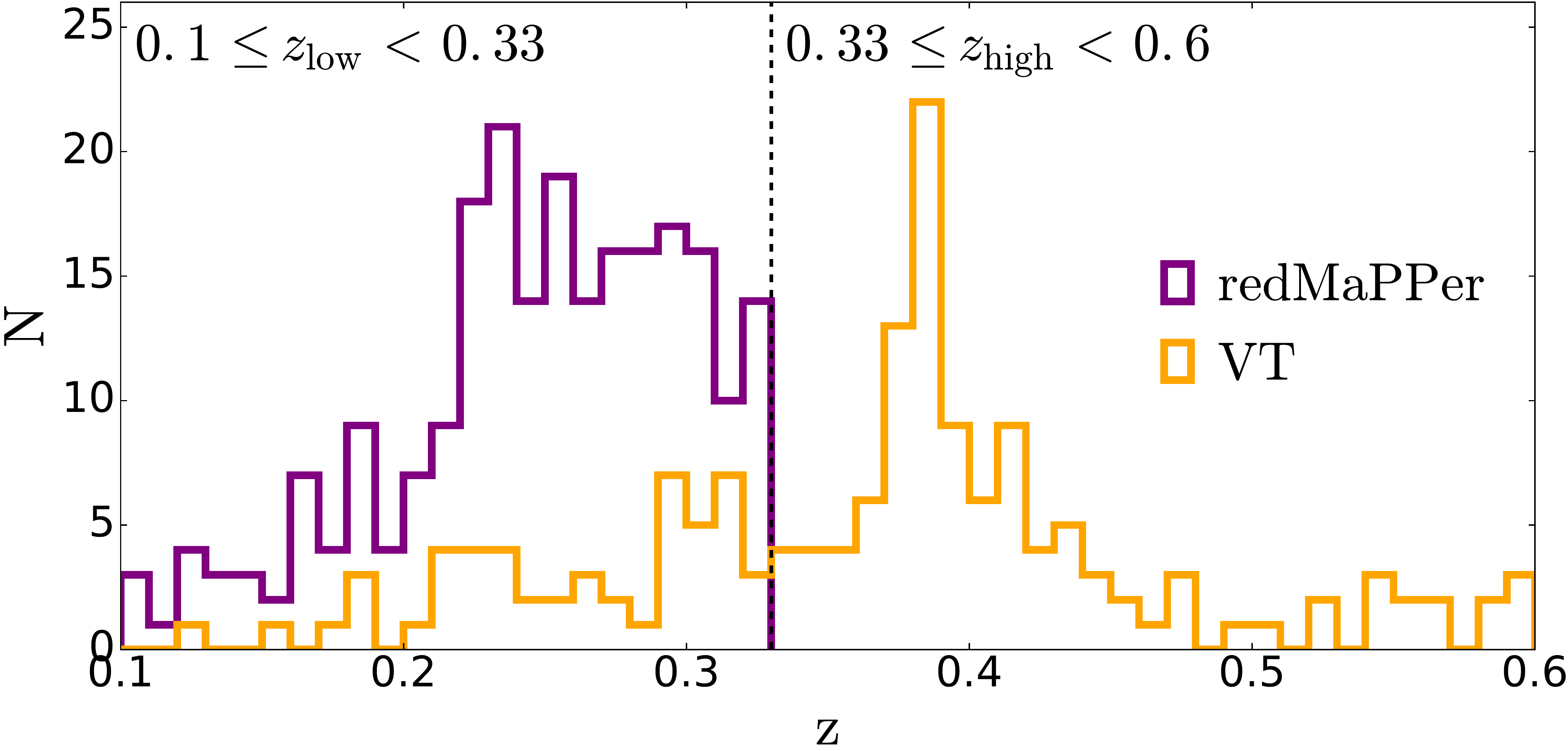}
 \caption{Redshift distributions of the redMaPPer (purple) and VT (orange) clusters used in our analysis. For our measurements we selected the redMaPPer sample in a low redshift bin ($0.1 \le z_{\mathrm{low}} <0.33$) and the VT sample in two redshift bins ($0.1 \le z_{\mathrm{low}} <0.33$ and $0.33\le z_{\mathrm{high}}<0.6$).}  
 \label{zdists}
\end{figure}

In Figure \ref{zdists} we show the redshift distributions for redMaPPer and VT clusters used in our stacked measurements highlighting the boundaries of the low and high $z$ bins. 
For the low redshift bin we follow \citeauthor{2017MNRAS.466.3103S} (\citeyear{2017MNRAS.466.3103S}, hereafter S17), and define $0.1 \le z_{\mathrm{low}} < 0.33$. We have 230 redMaPPer clusters at those redshifts, with $ 20 \leq \lambda \leq 128.7$.
The corresponding range of $\mu_{\star}$ for these clusters is $3.82 \times 10^{12} M_{\odot} \leq \mu_{\star} \leq 13.85  \times 10^{12}M_{\odot}$. 
For the VT sample we have 41 clusters in the low-redshift bin. We also consider a higher redshift bin, $0.33 \leq z_{\mathrm{high}} \leq 0.6$, for which there are 95 clusters in the catalog. 
The VT clusters in these two redshift bins lie within the range $1.47 \times 10^{12}M_{\odot} \leq \mu_{\star} \leq 16.53 \times 10^{12}M_{\odot}$. 

Inside each redshift bin, we separate the samples into four mass proxy bins, in such a way that we have a similar number of clusters in each bin. For the redMaPPer catalog we repeat this procedure twice, once for $\lambda$ and once for $\mu_{\star}$ (see Table \ref{tab_redmapper_lowz_binnig}). The stacking in $\lambda$ allows us to compare our mass-richness results with S17 and other %
measurements reported in the literature. The binning in $\mu_{\star}$ will enable us to compute the first mass-calibration of the redMaPPer cluster using this new mass proxy. 
Table \ref{tab_vt_lowzmidz_binnig} shows the $z$ and $\mu_{\star}$ bins for the VT catalog.

\begin{table}
	\centering
	\caption{Binning scheme and properties of the redMaPPer cluster sample. We use the same low redshift bin as S17, but for the binning in $\lambda$ we use a different scheme where we have a similar number of clusters in each of the four richness bins. Here $\mu_{\star}$ is given in units of $10^{12}M_{\odot}$.}
	\label{tab_redmapper_lowz_binnig}
	\begin{tabular}{cclcc} %
		\hline
		& Mean $z$ & $\lambda$ range & Mean $\lambda$ & No. of clusters\\
		\hline
         & 0.249 & $\left[20, 23.42\right)$ & 21.72 & 59\\
		 & 0.244 & $\left[23.42, 28.3\right)$ & 25.64 & 59\\
		 & 0.247 & $\left[28.3, 39.7\right)$ & 32.90 & 59\\
		 & 0.249 & $\left[39.7, 145\right)$ & 58.06 & 53\\
	\end{tabular}
    
	\begin{tabular}{cclcc} %
		\hline
        & Mean $z$ & $\mu_{\star}$ range & Mean $\mu_{\star}$ & No. of clusters\\
		\hline
         & 0.228 & $\left[0, 4.15\right)$ & 3.40 & 59\\
		 & 0.252 & $\left[4.15, 5.20\right)$ & 4.72 & 59\\
		 & 0.251 & $\left[5.20, 6.84\right)$ & 5.97 & 59\\
		 & 0.259 & $\left[6.84, 14\right)$ & 8.41 & 53\\
		\hline
	\end{tabular}    
\end{table}

\begin{table}
	\centering
	\caption{Binning scheme and properties of the VT cluster sample. We separate in two redshift bins and choose the 
$\mu_{\star}$ bins so as to have a similar number of clusters in each of the four bins. %
Here $\mu_{\star}$ is in units of $10^{12}M_{\odot}$.}
	\label{tab_vt_lowzmidz_binnig}
	\begin{tabular}{cclcc} %
		\hline
		$z$ range & Mean $z$ & $\mu_{\star}$ range & Mean $\mu_{\star}$ & No. of clusters\\
		\hline
		\multirow{4}{*}{[0.1, 0.33)} & 0.220 & $\left[0, 5.78\right)$ & 4.42 & 11\\
		 & 0.279 & $\left[5.78, 7.59\right)$ & 6.84 & 11\\
		 & 0.278 & $\left[7.59, 10.55\right)$ & 8.60 & 10\\
		 & 0.290 & $\left[10.55, 17\right)$ & 11.45 & 9\\
		\hline
		\multirow{4}{*}{[0.33, 0.6)} & 0.457 & $\left[0, 5.38\right)$ & 4.17 & 28\\
		 & 0.428 & $\left[5.38, 6.58\right)$ & 5.94 & 24\\
		 & 0.410 & $\left[6.58, 8.90\right)$ & 7.57 & 24\\
		 & 0.380 & $\left[8.90, 17\right)$ & 11.03 & 19\\
		\hline
	\end{tabular}    
\end{table}

\subsection{The stacked cluster profiles}
 
For any distribution of projected mass it is possible to show that the azimuthally averaged tangential shear $\gamma_t$ at a projected radius $R$ from the centre of the mass distribution \citep{1991ApJ...370....1M} is given by 
\begin{equation}
\gamma_t (R) = \frac{\Delta\Sigma}{\Sigma_{\mathrm{crit}}} \equiv \frac{\overline{\Sigma}(<R) - \langle \Sigma(R) \rangle}{\Sigma_{\mathrm{crit}}},
\label{deltasigmat}
\end{equation}
where $\Sigma(R)$ is the projected surface mass density at radius $R$, $\overline{\Sigma}(<R)$ is the mean value of $\Sigma$ within a disc of radius $R$, $\langle \Sigma(R) \rangle$ is the azimuthally averaged $\Sigma (R)$ within a ring of radius $R$ and $\Sigma_{\mathrm{crit}}$ is the critical surface mass density expressed in physical coordinates as 
\begin{equation}
\Sigma_{\mathrm{crit}} = \frac{c^2 D_s}{4 \pi G D_l D_{ls}},
\label{sigma_crit}
\end{equation}
where $D_l$ and $D_s$ are angular diameter distances from the observer to the lens and to the source, respectively, and $D_{ls}$ is the angular diameter distance between them.

From Equation \ref{deltasigmat} we can compute the %
surface density contrast $\Delta\Sigma$ over several lenses with similar physical properties (e.g. redshift, richness) to increase the lensing signal and reduce the effect of substructures, uncorrelated structures in the line of sight, shape noise and shape variations of individual halos. %

In practice we use the inverse variance weight $w$ from Lensfit to optimally weight shear measurements, accounting for shape measurement error and intrinsic scatter in galaxy ellipticity. Then, for a given lens $i$ and a given source $j$, the inverse variance weight for $\Delta\Sigma$ is derived for Equation \ref{deltasigmat} and expressed as $w_{ls, ij} = w_j \Sigma_{\mathrm{crit}, ij}^{-2}$. The quantity $w_{ls}$ is used to compute $\Delta\Sigma$ trough a weighted sum over all lens-source pairs    
\begin{equation}
\Delta\Sigma = \frac{\sum_{i=1}^{N_{l}} \sum_{j=1}^{N_s} w_{ls,ij} \times \gamma_{t,ij} \times \Sigma_{\mathrm{crit}, ij}}{\sum_{i=1}^{N_{l}} \sum_{j=1}^{N_s} w_{ls,ij} },
\end{equation}
where $N_l$ is the number of cluster lens and $N_s$ is the number of source galaxies.  

We compute $\Delta\Sigma$ in 20 logarithmically spaced radial bins from $R \sim 0.1\, h^{-1}$ Mpc to $R \sim 10\, h^{-1}$ Mpc. In \cite{2013MNRAS.429.2858M} it was pointed out that a multiplicative correction for the noise bias needs to be applied after stacking the shear. This correction can be computed from the multiplicative shear calibration factor $m(\nu_{\mathrm{SN}},r)$ provided by Lensfit. An often %
used expression for this correction \citep{2014MNRAS.437.2111V, 2015MNRAS.447..298H, 2017ApJ...840..104S, 2017MNRAS.467.3024L} is given by
\begin{equation}
1 + K (z_l) = \frac{\sum_{i=1}^{N_{l}} \sum_{j=1}^{N_{s}} w_{ls,ij} [1 + m(\nu_{\mathrm{SN}, ij}, r_{ij})] }{\sum_{i=1}^{N_{l}} \sum_{j=1}^{N_{s}} w_{ls,ij}},
\end{equation}
and the calibrated lensing signal is computed as
\begin{equation}
\langle \Delta\Sigma^{\mathrm{cal}} \rangle = \frac{\Delta\Sigma}{1 + K(z_l)}.
\label{SigmaCal}
\end{equation}

In order to reduce the dilution of the lensing signal due to uncertainties in the photo-zs that can cause some background sources to be placed as foreground sources and vice-versa, we impose that $z_s > z_l+0.1$ and $z_s > z_l + \sigma_{95}/2$ where $z_l$ is the lens redshift, $z_s$ is the source redshift and $\sigma_{95}$ is the 95 per cent confidence limit on the source redshift provided by BPZ. These cuts were validated by \cite{2017MNRAS.467.3024L}, who have found that the lensing signal is invariant over a range of lens-source separation cuts, suggesting that dilution caused by foreground or physically associated galaxies is not a large concern for CS82 weak lensing measurements (see their Appendix $A1$ for more details).         

{We compute the weak lensing signal $\Delta \Sigma$ from Eq. (\ref{SigmaCal}) in 20 logarithmic bins in the range $(0.1 - 10) h^{-1}$ Mpc}. As the errors on the weak-lensing signals are expected to be dominated by shape noise, we do not expect a noticeable covariance between adjacent radial bins and we treat them as independent in our analyses.
The error bars in our lensing signals are obtained by bootstrapping on the individual clusters with $N=100$ resamplings in each stack. \citet{vitorelli2017} have tested several bootstrap resampling values (e.g. $N=50$, 150, 200, 300) and found no significant variation of the error bars down to $R \lesssim 4$ Mpc. 

We computed the cross-component of the lensing signal ($\Delta\Sigma_{\times}$) and found no evidence of spurious correlations in the weak-lensing signals, i.e. the $\Delta\Sigma_{\times}$ measurements are consistent with zero. %
 
\subsection{Profile-fitting}

To model the average lensing signal around each lens and then obtain their mass estimates we use a model with two components: a perfectly centred dark matter halo profile and a miscentring term where the assumed centre does not correspond to the dynamical centre of the dark matter halo. For the first term we assume the clusters are well modeled by spherical Navarro--Frenk--White (NFW; \citealt{1996ApJ...462..563N}) haloes, on average, in which the 3-dimensional density profile is given by 
\begin{equation}
 \rho (r) = \frac{\delta_c \rho_{\mathrm{crit}}}{\frac{r}{r_s} \left( 1 + \frac{r}{r_s}\right)^2}, 
\end{equation}
where $r_s$ is the cluster scale radius, $\delta_{c}$ is the characteristic halo overdensity, $\rho_{\mathrm{crit}} = 3H^2(z)/8\pi G$ is the critical density of the universe at the lens redshift and $H(z)$ is the respective Hubble parameter.

{In this paper we use as cluster mass the mass $M_{200}$
contained within a radius $r_{200}$ where the mean mass density is 200 times the critical density of the universe. The scale radius is given by $r_s = r_{200}/c_{200}$, where $c$ is the so-called concentration parameter}. In our fitting procedure we follow \cite{2012A&A...545A..71V, 2015MNRAS.451.1460K} and use the concentration-mass scaling relation from \cite{2008MNRAS.390L..64D} given by    
\begin{equation}
c_{200} = 5.71 \times \left( \frac{M_{200} }{2 \times 10^{12} h^{-1}
} \right)^{-0.084} \times (1+z)^{-0.47}.
\end{equation}
\cite{1996A&A...313..697B, 1999astro.ph..8213O} provide %
an analytical expression for the projected NFW profile, $\Delta\Sigma_{\mathrm{NFW}}$ %
and we use a Python implementation\footnote{\url{https://github.com/joergdietrich/NFW}} of these results for our profile-fitting procedure.       

The central galaxy of a cluster is usually very bright but is not necessarily the BCG. For instance, \cite{2016ApJS..224....1R} pointed out that only $\sim 80-85$ per cent of the redMaPPer central galaxies are BCGs and \cite{2012MNRAS.426.2944Z} show that some BCGs present an offset from the centre of their host dark matter halo. This miscentring affects the observed shear profile \citep{2006MNRAS.373.1159Y,2007arXiv0709.1159J,2014MNRAS.439.3755F}. We follow the correction scheme presented in \cite{2007arXiv0709.1159J, 2015MNRAS.447.1304F, 2017MNRAS.466.3103S} to account for this effect. If the 2-dimensional offset in the lens plane is $R_s$, the azimuthal average of the profile is 
\begin{equation}
\Sigma_{\mathrm{misc}}(R) = \int_{0}^{\infty} dR_s P(R_s) \Sigma(R|R_s),  
\end{equation}
where
\begin{equation}
\Sigma(R|R_s) = \frac{1}{2\pi} \int_0^{2\pi} d\theta \Sigma\left(\sqrt{R^2+R_s^2+2RR_s\cos\theta}\right). 
\end{equation}
In other words, the angular integral of the profile $\Sigma (R)$ is shifted by $R_s$ from the centre. We also use a probability distribution for $R_s$ given by
\begin{equation}
P(R_s) = \frac{R_s}{\sigma_{\mathrm{off}}^2} \exp\left(-\frac{1}{2} \frac{R_s^2}{\sigma_{\mathrm{off}}^2}\right),
\end{equation}
which is an \textit{ansatz}, assuming the mismatching between the centre and $R_s$ follows a 2-dimensional Gaussian distribution. We use the Python implementation\footnote{\url{https://github.com/jesford/cluster-lensing}} of \cite{2016AJ....152..228F} to compute the miscentring term. The width of the miscentring distribution ($\sigma_{\mathrm{off}}$) is fixed as $0.4h^{-1}$ Mpc for simplicity. As noted in S17, this is an expected value for clusters with mass $\sim 10^{14} M_{\odot}$. %

Our complete theoretical modeling for $\Delta\Sigma$, considering the centred halo and miscentring terms, is given by
\begin{equation}
\Delta\Sigma^{\mathrm{theo}} = p_{\mathrm{cc}}\Delta\Sigma_{\mathrm{NFW}}+(1 - p_{\mathrm{cc}})\Delta\Sigma_{\mathrm{misc}}.
\label{delta_sigma_theor}
\end{equation}

In Table \ref{systematics} we present a summary of the systematics considered in this paper, both for obtaining the weak lensing signal $\Delta \Sigma$ and in the profile fitting.

\begin{table*}
\centering
\caption{Summary of the systematics we are take into account in the measurements of the lensing signal and in the profile-fitting. Note that since we apply a radial cut in innermost and outer range, following the same procedure as S17, our measurements are not affected by the central point mass and the 2-halo terms.}
\label{systematics}
\strutlongstacks{T}
\begin{tabular}{ll}
\hline
 Systematic: & Summary: \\
 \hline
 \hline
 Shear measurement & \Centerstack[l]{Apply additive calibration correction factor $c_2$ to $\epsilon_2$ component \\ Apply multiplicative shear calibration $m(\nu_{\mathrm{SN}} , r)$} 
\\ \hline
 Photometric redshifts & \Centerstack[l]{Remove $\mathrm{BPZ\_ODDS}\leq 0.5$ to reduce systematic errors due to catastrophic outliers \\ Apply $z_s > z_l + 0.1$ and $z_s > z_l + \frac{\sigma_{95} }{2}$} 
\\ \hline 
 miscentring & \Centerstack[l]{Apply same correction as \citealt{2006MNRAS.373.1159Y, 2007arXiv0709.1159J, 2017ApJ...840..104S} } \\ \hline  
\end{tabular}
\end{table*}

In addition to the contribution from single (centred and miscentred) cluster halos, a variety of studies in the literature have pointed out the need to consider other terms to better model the measured profile. These often include a \textit{point mass} term for a possible stellar-mass contribution of the central galaxies and a so-called \textit{2-halo} term due to neighbouring halos (i.e., due to the large-scale structure of the Universe). In this work, we avoid these two contributions as we are only interested in measuring $M_{200}$ and we do not have enough precision to fit for many free parameters in each mass-proxy bin. For this sake, we perform the model-fitting in a restricted radial range. We follow S17 and use $R_{\min}=0.3\,h^{-1}$ Mpc as the inner radius limit to avoid problems with the selection of background galaxies and the increased scatter due to the low sky area, and also to reduce the effects of the point mass contribution \citep[see also][]{2010MNRAS.405.2078M}. We define a richness-dependent outer limit in the range $R_{\max} \simeq (2.5 - 3.5)\,h^{-1}$ Mpc to avoid the 2-halo contribution. S17 shows that the results are insensitive to the specific values of $R_{\max}$ for a wide range of values.

Finally, for each sample in the radial range mentioned above, we perform the profile-fitting via Bayesian formalism and Monte Carlo Markov Chain (MCMC) method to compute the \textit{posterior} distribution $Pr(M_{200},p_{\mathrm{cc}} | \Delta\Sigma^{\mathrm{obs}})$ and then obtain the best estimate for the cluster mass. Following \cite{vitorelli2017}, we use a flat prior for the mass ($10^{12} h^{-1} M_{\odot} <M_{200}< 10^{15} h^{-1} M_{\odot}$) and a Gaussian prior on the miscentring term, $\mathcal{N} (p_{\mathrm{cc}};\overline{P_{\mathrm{cen}}}, \sigma_{P_{\mathrm{cen}}})$, for $0<p_{\mathrm{cc}}<1$, where $\overline{P_{\mathrm{cen}}}$ and $\sigma_{P_{\mathrm{cen}}}$ are the mean and standard deviation of the highest centring probabilities $P_{\rm cen}$. We use the same modeling approach for $P_{\mathrm{cen}}^{\star}$, in both the redMaPPer and VT catalogs.    

\begin{figure*}    
\includegraphics[width=\textwidth]{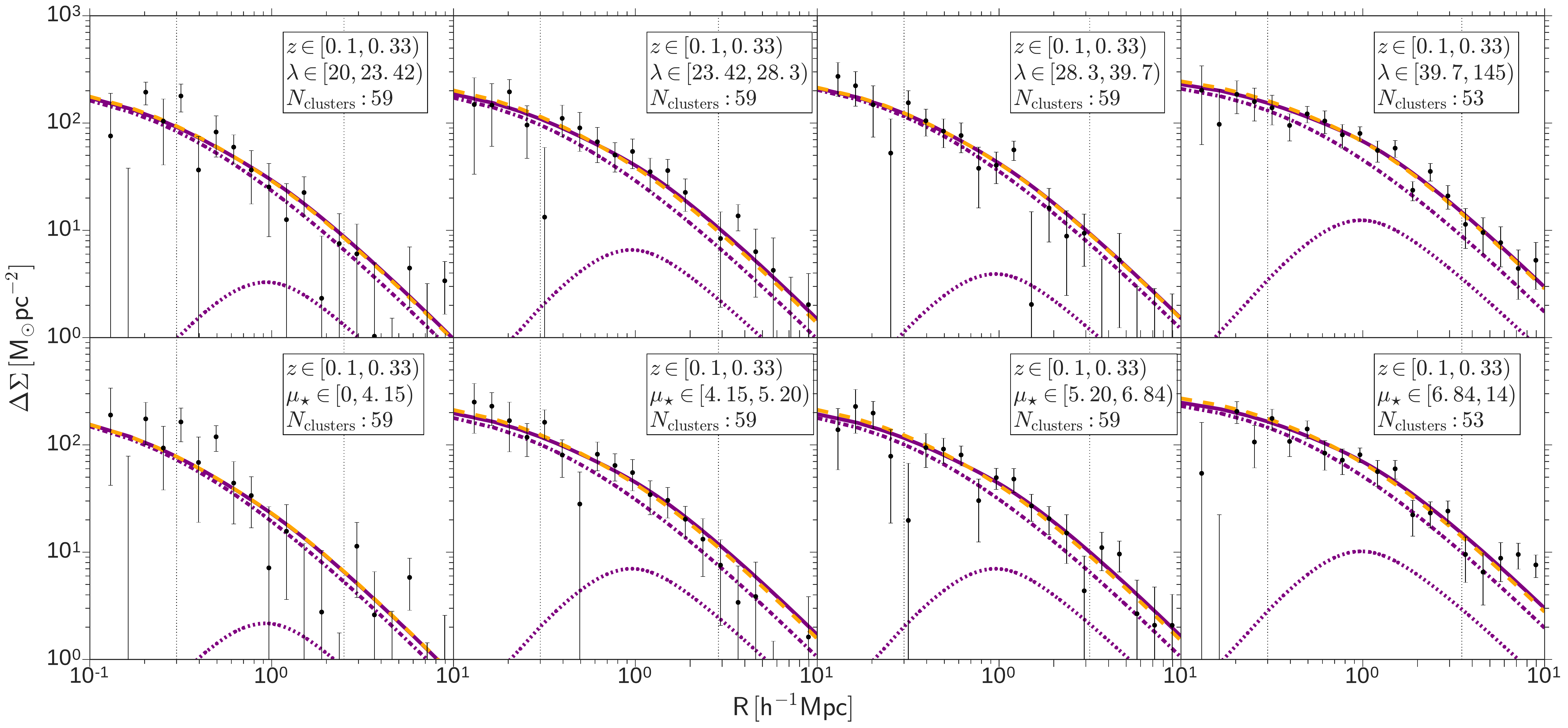}
\caption{The $\Delta\Sigma$ measurements and the profile-fitting results for the stacked redMaPPer clusters in the low redshift interval in four observable mass proxy bins. In the top panel we show the results for the binning using $\lambda$ and in the bottom panel the results for binning in $\mu_{\star}$ (in units of $10^{12}M_{\odot}$). 
{The fit to the models is performed for the radial bins within the two vertical dotted lines.}
The \textit{purple solid} line shows the best-fit results for a combination of NFW and miscentring term, using the information of $P_{\mathrm{cen}}$ as a \textit{prior} for the miscentring offset. The \textit{orange dashed} line shows the best-fit using $P_{\mathrm{cen}}^{\star}$ as the information for the \textit{prior} when performing the fit.
{The dashed-dotted and dotted lines show the contribution of the two terms to the best fit profile: the centred NFW profile 
$p_{\mathrm{cc}}\Delta\Sigma_{\mathrm{NFW}}$ (\textit{purple dashed-dotted})} and the %
miscentring term $(1 - p_{\mathrm{cc}})\Delta\Sigma_{\mathrm{misc}}$ (\textit{purple dotted}).}
\label{DSfit_rm_lambda_mustar_bins}
\end{figure*}

In Figure \ref{DSfit_rm_lambda_mustar_bins} we show the weak lensing profiles for the redMaPPer clusters. We present the measured signal (\textit{black} dots) and the best fits using $P_{\mathrm{cen}}$ in the Gaussian prior for miscentring (\textit{purple solid} line) and using $P_{\mathrm{cen}}^{\star}$ in the prior (\textit{orange dashed} line). We also show the centred halo contribution (\textit{purple dotted-dashed} line) and the miscentring term (\textit{purple dotted} line) from Eq. (\ref{delta_sigma_theor}) as computed in the $P_{\mathrm{cen}}$ prior case. The dotted vertical lines correspond to $R_{\max}$ and $R_{\min}$, which define the range were the fit is performed. We show the low redshift sample in bins of $\lambda$ (in the top panel) and $\mu_{\star}$ (in the bottom).

\begin{figure*}    
\includegraphics[width=\textwidth]{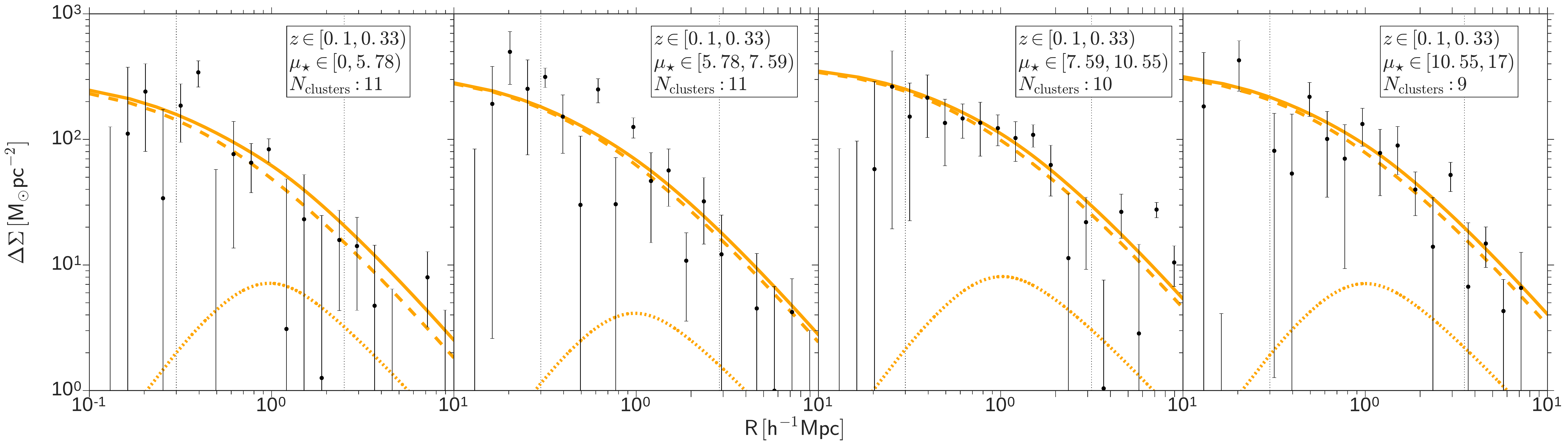}
\caption{The $\Delta\Sigma$ measurements and the profile-fitting results for the stacked VT clusters in the low redshift interval in four bins of $\mu_{\star}$ (in units of $10^{12}M_{\odot}$). The fit is performed for the radial bins within the two vertical dotted lines. The \textit{orange solid} line shows the best-fit using  $P_{\mathrm{cen}}^{\star}$ in the \textit{prior} for the miscentring offset in the fit. {The dashed and dotted lines show the contribution of the two terms to the best fit profile: the centred NFW profile (\textit{orange dashed}) and the} miscentring term (\textit{orange dotted}).} 
\label{DSfit_vt_mustar_bins_lowz}
\end{figure*}

\begin{figure*}    
\includegraphics[width=\textwidth]{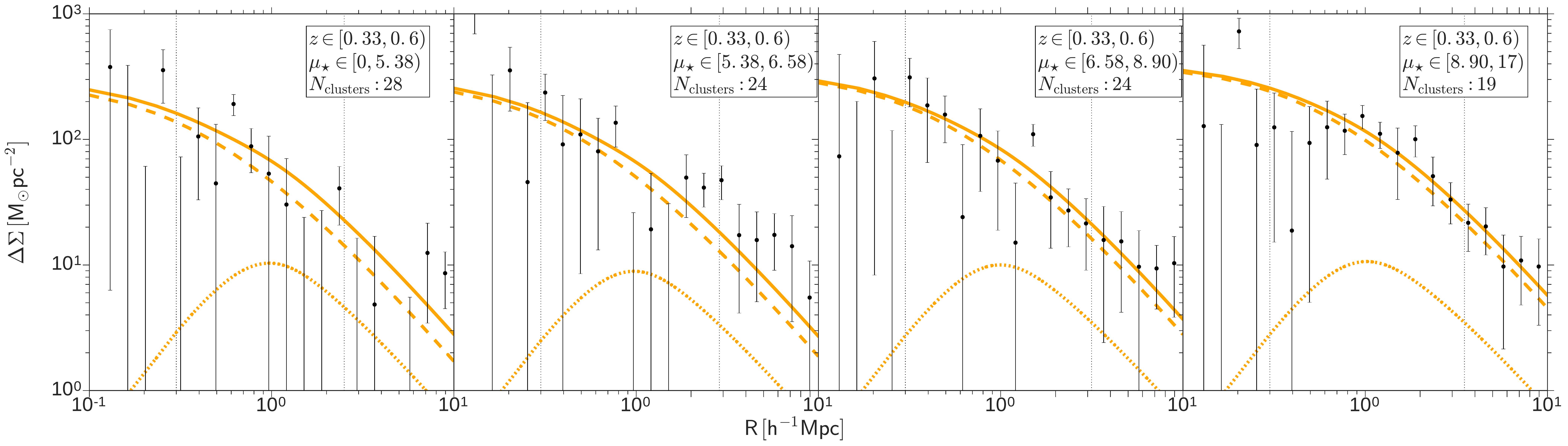}
\caption{Same as the previous figure, but for the interval $0.33 \le z_{\mathrm{high}} <0.6$ of the VT clusters in bins of $\mu_{\star}$ (intervals in units of $10^{12}M_{\odot}$).}
\label{DSfit_vt_mustar_bins_midz}
\end{figure*}

We see from Figure \ref{DSfit_rm_lambda_mustar_bins} that the best fit results using $P_{\mathrm{cen}}$ and $P_{\mathrm{cen}}^{\star}$ are very similar, validating the use of $P_{\mathrm{cen}}^{\star}$ for the miscentring correction, and in particular its application to the VT clusters. In Figures \ref{DSfit_vt_mustar_bins_lowz} and \ref{DSfit_vt_mustar_bins_midz} we show the profile-fitting results for the VT clusters in the low and high redshift samples in bins of $\mu_{\star}$. The best-fit values of the two parameters for all cases considered here are presented in Table \ref{tab_fitting_results}. In our analyses we use $M_{200}$ relative to critical matter density (hereafter $M_{200c}$) of the Universe, however, to enable the comparison with other works in the literature, it is useful to express the results in terms of $M_{200}$ relative to the mean density ($M_{200m}$). To convert from $M_{200c}$ to $M_{200m}$ we use the Colossus code\footnote{\url{https://bitbucket.org/bdiemer/colossus}} \citep{2015ascl.soft01016D}. In Table \ref{tab_fitting_results} we show the results in terms of both mass definitions.
 
\begin{table*}
	\centering
\caption{Best-fit results for redMaPPer clusters in Figure \ref{DSfit_rm_lambda_mustar_bins} and for VT clusters in Figures \ref{DSfit_vt_mustar_bins_lowz} and \ref{DSfit_vt_mustar_bins_midz}. In the fitting we use a concentration-mass relation from \protect\cite{2008MNRAS.390L..64D} to fix $c_{200}$ and we fix the width of miscentring distribution as $\sigma_{\mathrm{off}}=0.4\,h^{-1}$ Mpc. Our final model has just two free parameters, the mass $M_{200}$ (computed using the critical density and converted to the mean density with Colossus) and the fraction of clusters that is correctly centred $p_{\mathrm{cc}}$. For the redMaPPer clusters, we use the mean and standard deviation ($\sigma$) of $P_{\mathrm{cen}}$ in a Gaussian prior for $p_{\mathrm{cc}}$, while for the VT clusters we use the mean and $\sigma$ of $P_{\mathrm{cen}}^{\star}$ for the Gaussian prior. The values of $\mu_{\star}$ that define each stack are given in units of $10^{12}M_{\odot}$.}
	\label{tab_fitting_results}    
	\begin{tabular}{lcccc} %
        \hline
		  & z & $M_{200c} \,\, (10^{14} h^{-1} M_{\odot})$ & $M_{200m} \,\, (10^{14} h^{-1} M_{\odot})$ & $p_{\mathrm{cc}}$ \\
		\hline
        \multicolumn{4}{|l|}{redMaPPer} \\
		\hline       
		$20 \leq \lambda < 23.42$ & \multirow{4}{*}{$0.1 \leq z < 0.33$} & $0.83 \pm 0.23$ & $1.08 \pm 0.30$ & $0.83 \pm 0.11$\\
		$23.42 \leq \lambda < 28.3$ &  & $1.30 \pm 0.38$ & $1.71 \pm 0.49$ & $0.75 \pm 0.13$\\
		$28.3 \leq \lambda < 39.7$ &  & $1.30 \pm 0.27$ & $1.71 \pm 0.35$ & $0.86 \pm 0.10$\\
		$39.7 \leq \lambda < 145$ &  & $2.90 \pm 0.55$ & $3.84 \pm 0.71$ & $0.71 \pm 0.12$\\
                \hline
		$0 \leq \mu_{\star} < 4.15$ & \multirow{4}{*}{$0.1 \leq z < 0.33$} & $0.59 \pm 0.19$ & $0.77 \pm 0.25$ & $0.86 \pm 0.10$\\
		$4.15 \leq \mu_{\star} < 5.20$ &  & $1.60 \pm 0.42$ & $2.10 \pm 0.54$ & $0.75 \pm 0.13$\\
		$5.20 \leq \mu_{\star} < 6.84$ &  & $1.50 \pm 0.36$ & $1.97 \pm 0.47$ & $0.75 \pm 0.13$\\
		$6.84 \leq \mu_{\star} < 14$ &  & $2.90 \pm 0.52$ & $3.82 \pm 0.67$ & $0.77 \pm 0.11$\\
		\hline
        \hline
        \multicolumn{4}{|l|}{VT} \\
		\hline
		$0 \leq \mu_{\star} < 5.78$ & \multirow{4}{*}{$0.1 \leq z < 0.33$} & $2.40 \pm 0.65$ & $3.20 \pm 0.85$ & $0.82 \pm 0.10$\\
		$5.78 \leq \mu_{\star} < 7.59$ &  & $2.50 \pm 0.49$ & $3.27 \pm 0.63$ & $0.91 \pm 0.04$\\
		$7.59 \leq \mu_{\star} < 10.55$ &  & $5.20 \pm 1.10$ & $6.86 \pm 1.42$ & $0.89 \pm 0.03$\\
		$10.55 \leq \mu_{\star} < 17$ &  & $3.80 \pm 0.92$ & $4.97 \pm 1.18$ & $0.88 \pm 0.04$\\
        \hline
		$0 \leq \mu_{\star} < 5.38$ & \multirow{4}{*}{$0.33 \leq z < 0.6$} & $2.50 \pm 0.92$ & $3.09 \pm 1.13$ & $0.75 \pm 0.12$\\
		$5.38 \leq \mu_{\star} < 6.58$ &  & $2.40 \pm 0.90$ & $2.99 \pm 1.11$ & $0.79 \pm 0.09$\\
		$6.58 \leq \mu_{\star} < 8.90$ &  & $3.30 \pm 0.81$ & $4.15 \pm 1.00$ & $0.82 \pm 0.07$\\
		$8.90 \leq \mu_{\star} < 17$ &  & $5.50 \pm 1.00$ & $7.01 \pm 1.25$ & $0.86 \pm 0.04$\\
		\hline
	\end{tabular}    
\end{table*}

\section{Results}
\label{results}

From the weak lensing masses in Table \ref{tab_fitting_results} we obtain a mass calibration for redMaPPer clusters and compare with the current results from the literature. We then apply the same methodology to obtain the mass-observable scaling relation for the new mass proxy $\mu_{\star}$, both for the redMaPPer and VT clusters.  

In this work, the mass-richness relation for the redMaPPer mass proxy $\lambda$ is given by the power law expression 
\begin{equation}
\langle M_{200} | \lambda \rangle = M_0 \left(\frac{\lambda}{\lambda_0}\right)^{\alpha},
\label{massrich}
\end{equation}
where $\lambda_0$ is a fixed pivot richness and the normalization $M_0$ and the slope $\alpha$ are the free parameters. 

For the new mass proxy we fit a power-law relation to the mass obtained in the $\mu_{\star}$ bins akin to Equation (\ref{massrich}):
\begin{equation}
\langle M_{200} | \mu_{\star} \rangle = M_0 \left(\frac{\mu_{\star}}{\mu_{\star}^0}\right)^{\alpha},
\label{massstellar}
\end{equation}
where the pivot value $\mu_{\star}^0$ is chose as the median value of the proxy in each sample. 

\subsection{redMaPPer mass-richness relation}
\label{rm-results}

To validate our mass estimates we make a comparison with S17, which uses the same redMaPPer catalogue in the same low redshift bin to compute a mass-richness relation. However, the analysis in S17 is not limited to the SDSS Stripe 82 region, which implies that they have more statistics than us. On the other hand, our shape measurements are made in better quality images than SDSS and using the state-of-the-art code Lensfit, which enables us to have a good SNR for our lensing signal to make this comparison. 

\begin{figure}
\centering
 \includegraphics[width=\columnwidth]{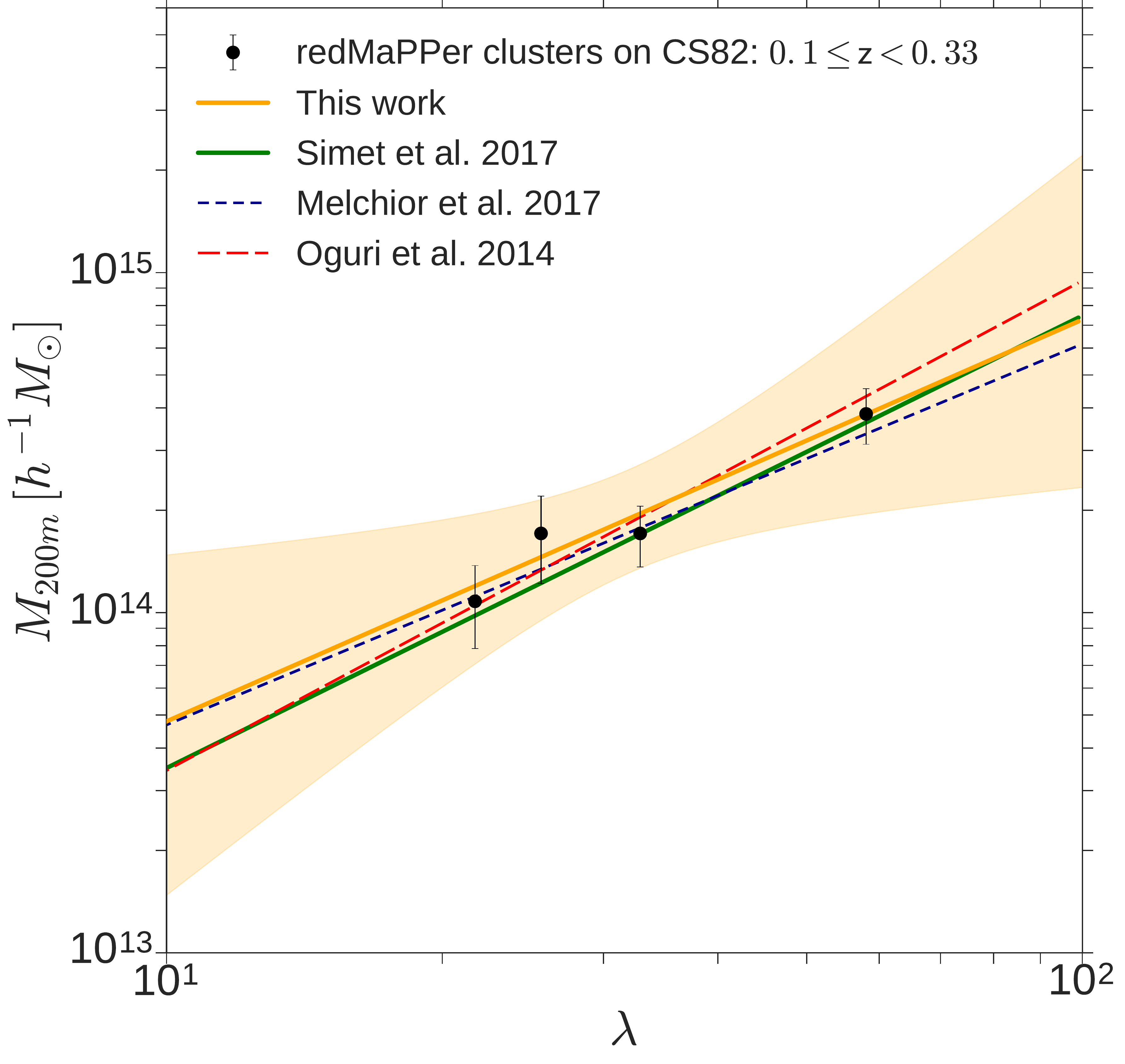}
\caption{Comparison of the mass-richness relations of redMaPPer clusters in the $z_{\mathrm{low}}$ interval. The adopted $z_{\mathrm{low}}$ interval is the same for S17 (\textit{green solid} line) and this work (\textit{orange solid} line). \citet[][\textit{blue dashed} line]{2016arXiv161006890M} work in the range $0.2 \leq z < 0.4$ while \citet[][\textit{red double-dashed} line]{2014MNRAS.444..147O} use the range $0.1 < z < 0.3$ for its low redshift interval. 
We show the $2\sigma$ confidence intervals (\textit{orange shaded} region) for the cluster mass $M_{200m}$ as a function of the richness $\lambda$ from this work. The value of the normalizations and slopes are shown in Table \protect\ref{massrichlit2}.}
\label{massrich_rm_lambda_lowz}
\end{figure}

In Figure \ref{massrich_rm_lambda_lowz} we show our best-fit $M_{200m}$ versus $\lambda$ relation (\textit{orange solid} line) and its $2\sigma$ confidence intervals (\textit{orange shaded} regions). We show, for  comparison, the S17 mass-richness relation (\textit{green solid} line). Using the same pivot richness as S17, $\lambda_0=40$, we find $M_0=(2.46\pm 0.44)\times10^{14}h^{-1}M_{\odot}$ and $\alpha=1.18\pm 0.38$ while they have obtained $M_0=(2.21\pm0.22)\times10^{14}h^{-1}M_{\odot}$ and $\alpha=1.33_{-0.10}^{+0.09}$. Additionally, we present the mass-richness relation obtained by \cite[][\textit{blue dashed} line]{2016arXiv161006890M} for clusters identified with redMaPPer in the DES Science Verification data, with shears measured on that same data, in a similar low redshift bin ($0.2<z_{\mathrm{low}}<0.4$). Their results, converted to our units and pivot $\lambda_0=40$, are $M_0=(2.21\pm0.35)\times10^{14}h^{-1}M_{\odot}$ and $\alpha=1.12\pm 0.20$. We also compare our results to the mass-richness relation for the red sequence based CAMIRA code of \cite{2014MNRAS.444..147O}. The CAMIRA code was applied to the same SDSS DR8 data and has its own richness estimator, $\hat{N}_{\mathrm{cor}}$. In order to convert their result to our units, we first performed a cylindrical match between our sample and their catalog to find the mean relation between $\hat{N}_{\mathrm{cor}}$ and $\lambda$. Our cylindrical match uses a search radius of 1 arcmin and $\Delta z=0.05$. We found 339 matched clusters from which we derived the CAMIRA-redMaPPer richness scaling relation $\hat{N}_{\mathrm{cor}}=A\lambda$ with $A=0.819\pm 0.009$. The mass calibration for CAMIRA is obtained for $M_{200\mathrm{vir}}$, which we convert to $M_{200m}$ using Colossus, and we converted their calibration to the pivot $\lambda_0=40$ as well. We find that their converted results are $M_0=(2.53\pm 0.30)\times10^{14}h^{-1}M_{\odot}$ and $\alpha=1.44\pm0.27$ (\textit{red double-dashed} line). These results are summarized in Table \ref{massrichlit2}.

Despite using different data and slightly different approaches, we see that our mass measurements are in excellent agreement with those results from the literature, which validates our methodology to obtain average mass estimates from the stacked weak lensing signal.

\begin{table}
\centering
\caption{Comparison of the redMaPPer mass-richness relation in the $z_{\mathrm{low}}$ bin with three recent results from the literature. The normalization $M_0$ from \protect\cite{2016arXiv161006890M} is converted to our units. We also have to convert $M_0$ from \protect\cite{2014MNRAS.444..147O} to our units and find a relation between their richness $\hat{N}_{\mathrm{cor}}$ and $\lambda$. All calibrations are computed or converted to the pivot $\lambda_0=40$.}
\begin{tabular}{lcc}
          &  $M_0 (10^{14}h^{-1}M_{\odot} ) $ & $\alpha$ \\
\hline
This work & $2.46\pm0.44$ & $1.18\pm0.38$ \\
Simet et al. 2017 & $2.21\pm0.22$ & $1.33^{+0.09}_{-0.10}$ \\
Melchior et al. 2017 & $2.21\pm0.35$ & $1.12\pm0.20$ \\
Oguri et al. 2014 & $2.53\pm0.30$ & $1.44\pm0.27$ \\
\end{tabular}
\label{massrichlit2}
\end{table}

\begin{figure}
 \includegraphics[width=\columnwidth]{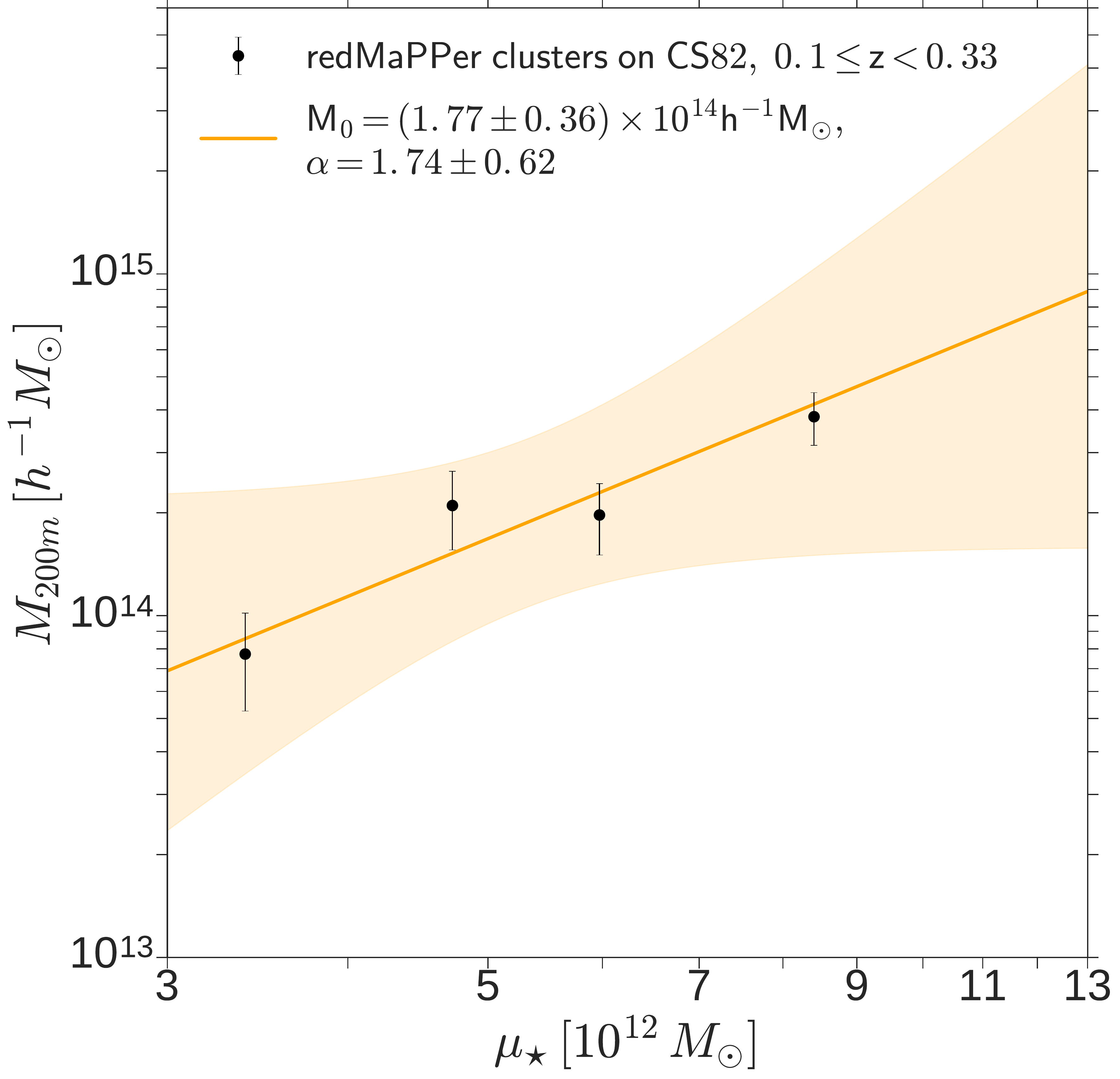}
 \caption{Mass-calibration with $2\sigma$ confidence intervals for redMaPPer clusters binned in $\mu_{\star}$ in the $z_{\mathrm{low}}$ interval. For the mass estimates we apply the miscentring correction. In the mass--$\mu_{\star}$ relation we adopt the median of $\mu_{\star}$ as the mass proxy pivot, $\mu_{\star}^0 = 5.16 \times 10^{12} M_{\odot}$.}%
\label{massrich_rm_mustar_lowz}
\end{figure}

As mentioned, we also computed $\mu_{\star}$ for the redMaPPer clusters. We fit the power-law relation of Equation (\ref{massstellar}) with pivot value $\mu_{\star}^0 = 5.16 \times 10^{12} M_{\odot}$. We find $M_0=(1.77\pm0.36)\times10^{14}h^{-1}M_{\odot}$ and $\alpha=1.74\pm0.62$. In Figure \ref{massrich_rm_mustar_lowz} we show the best fit $ M_{200m} \times \mu_{\star}$ relation (\textit{orange solid} line) and its $2\sigma$ confidence intervals (\textit{orange shaded} region) for the $z_{\mathrm{low}}$ interval. 

\subsection{VT -- \texorpdfstring{$\mu_\star$}{m} mass-calibration}

\begin{figure}
 \includegraphics[width=\columnwidth]{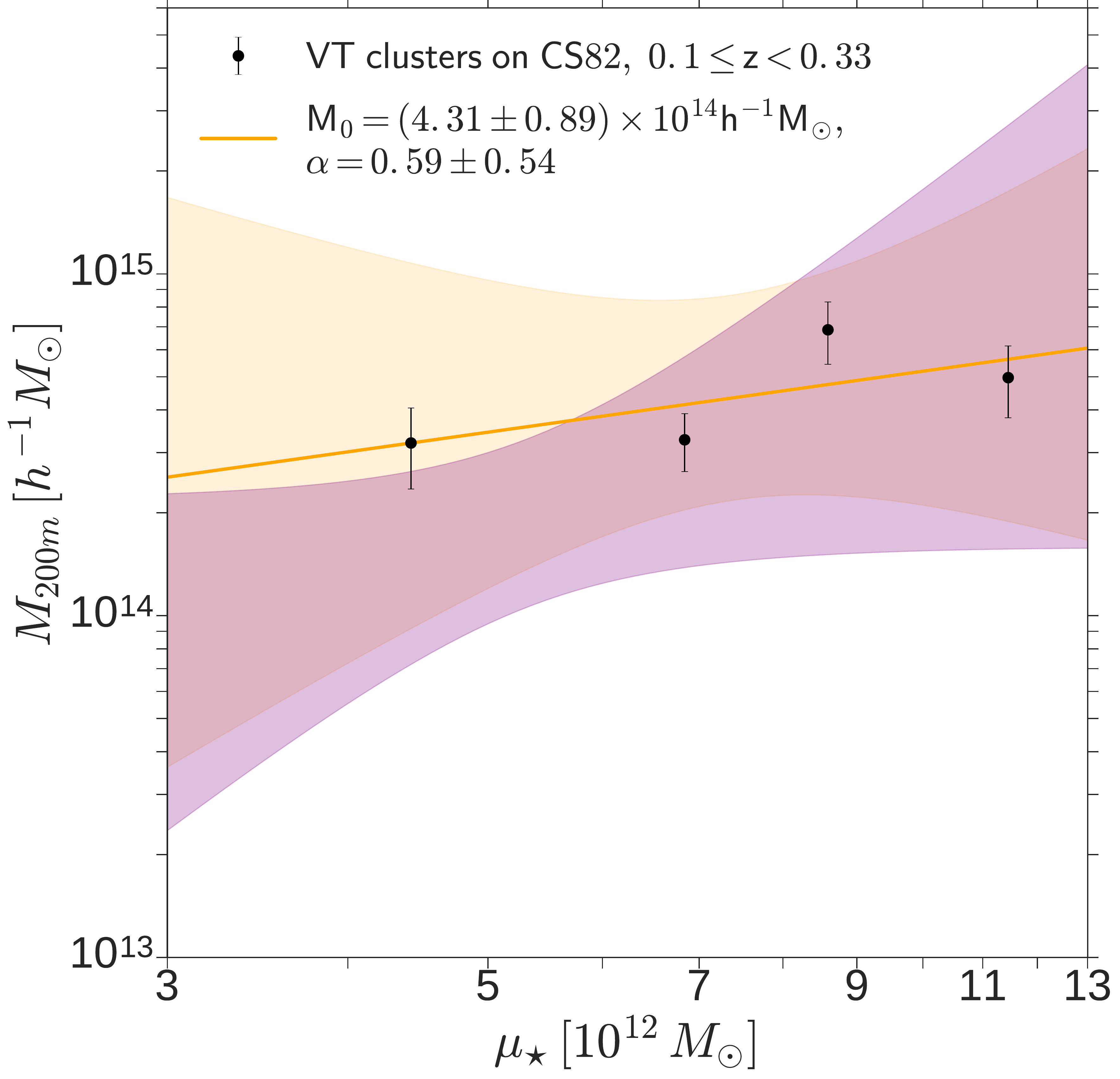}
 \caption{Mass-calibration with $2\sigma$ confidence intervals (\textit{orange shaded} regions) for VT clusters binned in $\mu_{\star}$ in the $z_{\mathrm{low}}$ interval. Miscentring corrections were applied in the mass estimates. In the mass-richness relation the pivot is $\mu_{\star}^0 = 7.30 \times 10^{12} M_{\odot}$. For comparison, we also present the $2\sigma$ confidence intervals (\textit{purple shaded} regions) for the redMaPPer $z_{\mathrm{low}}$ clusters.}%
 \label{massrich_vt_mustar_lowz}
\end{figure}

In Figure \ref{massrich_vt_mustar_lowz} we show $M_{200m} \times \mu_{\star}$ for VT clusters in the $z_{\mathrm{low}}$ interval, following the same approach we used to calibrate the mass as a function of $\mu_{\star}$ in the redMaPPer cluster sample. The \textit{orange solid} line is the best-fit result and the \textit{orange shaded} regions are the $2\sigma$ confidence intervals for this VT sample. The pivot is $\mu_{\star}^0 = 7.30 \times 10^{12} M_{\odot}$ and we find $M_0=(4.31\pm0.89)\times10^{14}h^{-1}M_{\odot}$ and $\alpha=0.59\pm0.54$. For comparison, we show as \textit{purple shaded} regions the same $2\sigma$ confidence intervals obtained for the redMaPPer clusters shown in Figure \ref{massrich_rm_mustar_lowz}. 
We see a good agreement at this confidence level, despite the fact that the cluster samples are significantly different. Actually if we consider the VT and redMaPPer data points altogether, i.e. if we combine the VT $\mu_{\star}$ bins and corresponding masses and the redMaPPer $\mu_{\star}$ bins and respective masses, we obtain a power-law fit as good as the one for the VT points only. In other words the redMaPPer mass-$\mu_{\star}$ relation is compatible to the VT one.

\begin{figure}
 \includegraphics[width=\columnwidth]{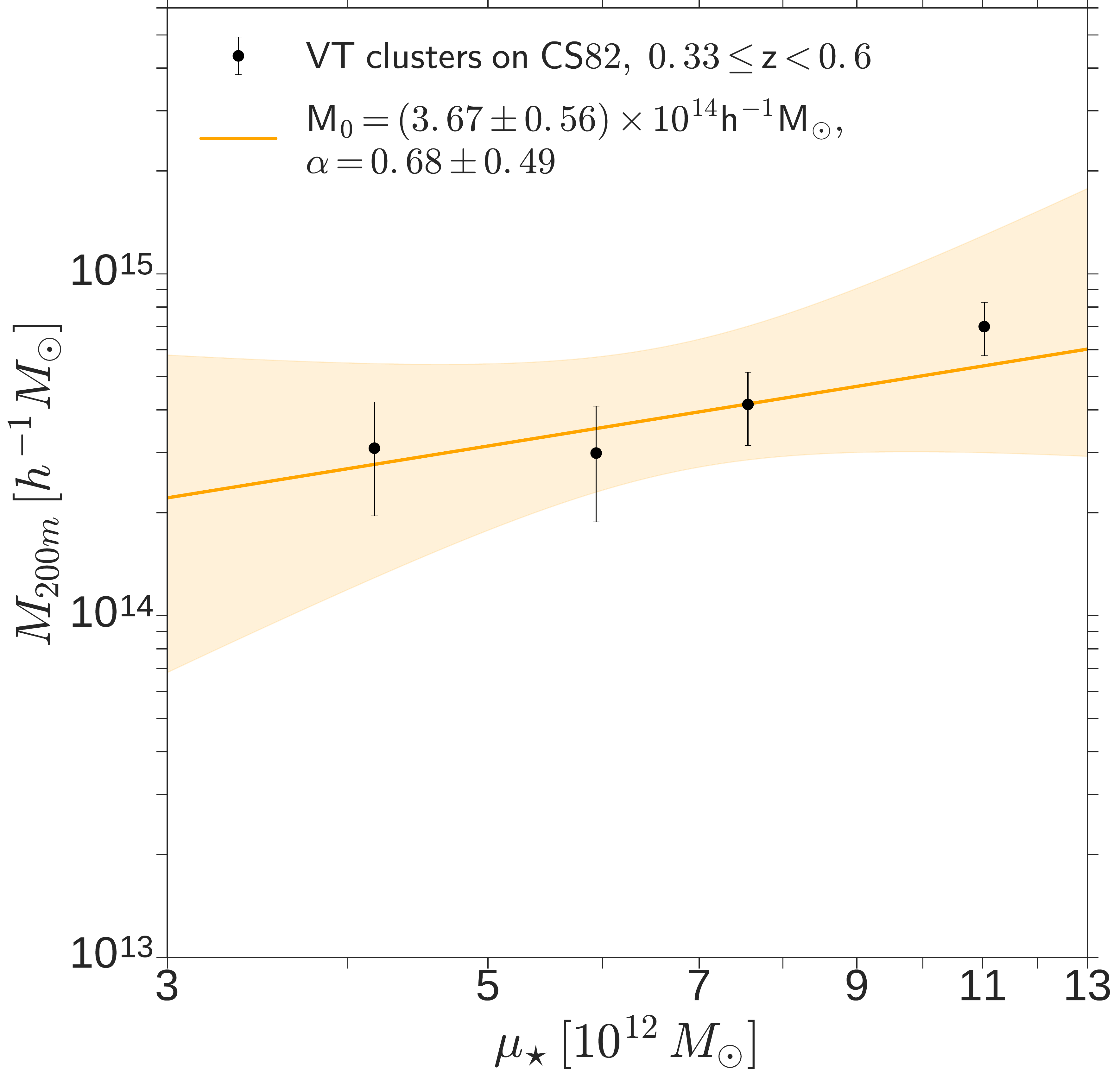}
 \caption{Same as previous figure but for the $z_{\mathrm{high}}$ interval and pivot $\mu_{\star}^0 = 6.30 \times 10^{12} M_{\odot}$.} 
 \label{massrich_vt_mustar_midz}
\end{figure}

We present the mass-calibration results for the $z_{\mathrm{high}}$ interval of VT clusters in Figure \ref{massrich_vt_mustar_midz}. The \textit{orange solid} line and \textit{orange shaded} regions are the best-fit and the $2\sigma$ confidence intervals, respectively. We have used a pivot $\mu_{\star}^0 = 6.30 \times 10^{12} M_{\odot}$ and find $M_0=(3.67\pm0.56)\times10^{14}h^{-1}M_{\odot}$ and $\alpha=0.68\pm0.49$. As previously mentioned, we were able to extend our analysis of the VT sample to the higher redshift range $0.33<z<0.6$ because the VT clusters were identified in the SDSS co-add data, which is deeper than SDSS single epoch data used to identify the redMaPPer sample. {In addition, the CS82 shear catalog is still reliable for lenses at these redshifts}. The results of the all mass-$\mu_{\star}$ calibrations are summarized in Table \ref{massmustarfinal}.  

\begin{table}
\centering
\caption{Summary of mass-$\mu_{\star}$ calibration for redMaPPer (RM) and VT clusters obtained from this work. We present the normalization $M_0$ and slope $\alpha$ values by fitting the Equation \ref{massstellar} as well the proxy pivot $\mu_{\star}^0$ adopted for each sample, i.e. RM clusters at $0.1 \leq z_{\mathrm{low}}<0.33$ and VT clusters in the same RM low redshift bin as well as a high redshift bin $0.33 \leq z_{\mathrm{high}}<0.6$.}
\begin{tabular}{cccc}
Sample & $\mu_{\star}^0 (10^{12}M_{\odot} )$ & $M_0 (10^{14}h^{-1}M_{\odot} )$ & $\alpha$ \\
\hline
RM $z_{\mathrm{low}}$ & 5.16 & $1.77\pm0.36$ & $1.74\pm0.62$ \\
VT $z_{\mathrm{low}}$ & 7.30 & $4.31\pm0.89$ & $0.59\pm0.54$ \\
VT $z_{\mathrm{high}}$ & 6.30 & $3.67\pm0.56$ & $0.68\pm0.49$ \\
\end{tabular}
\label{massmustarfinal}
\end{table}

\section{Discussion}\label{conclusions}
We perform a weak lensing mass calibration of $\mu_{\star}$, a cluster mass proxy that includes information about galaxies regardless of their colour. Unlike the empirically determined red sequence--based mass proxies, $\mu_{\star}$ is physically motivated:  the stellar mass inside a dark matter halo can be expected to trace the dark matter halo mass. Furthermore, it turns out that the stellar mass is a relatively robust observable \citep[see][and references therein]{2013ARA&A..51..393C} and independent of the history of the formation of the red sequence. The redshift at which the red sequence forms in clusters is not currently known, and at high enough redshifts redMaPPer will become increasingly incomplete in terms of finding %
dark matter halos. Additionally, stellar masses are easier to model in simulations than the red sequence 
\citep[e.g.,][]{2017ApJ...836..120R}.
 
It is natural to use a well-studied sample of clusters in the development of a new mass proxy and to use a well-studied mass proxy to validate our methodology. We have measured the redMaPPer $\lambda$-mass scaling relation and showed results consistent with similar scaling relations reported in the literature. 
We then performed the scaling relation measurement on the same redMaPPer clusters binning on $\mu_{\star}$ instead of $\lambda$. The most direct comparison between the two scaling relations is made at the pivot point: the slope and the mass at the pivot point are consistent between the $\lambda$ and $\mu_{\star}$ proxies. 

Since we applied the methodology on the same clusters in measuring both scaling relations, our results can be directly interpreted. Imagine a scenario in which all cluster members are in the red sequence. There would be a maximal correlation between $\lambda$ and $\mu_{\star}$ as all red galaxies have very similar mass-to-light ratios. The scaling relations would, therefore, be nearly identical. If we change the scenario to include blue galaxies and compute a $\lambda$-like proxy, the slope of the $\lambda$-like proxy with mass would be shallower because the luminosity of the blue galaxies is most often driven by single star formation events and the high luminosity of young massive stars, and large numbers of low luminosity galaxies would be pushed above the threshold. If the $\mu_{\star}$ proxy were similarly affected by blue galaxies our measured slope would be shallow. The fact that our measurements of the scaling relations in redMaPPer are so close to each other indicate that the stellar mass in these systems is tracing dark matter mass with not much worse scatter than $\lambda$. In low $z$ clusters it is known that nearly all members are red and therefore our results are not surprising here. At high redshift, however, this is not true. A red-sequence selected high $z$ sample might show a significant difference between $\lambda$ and $\mu_{\star}$ mass calibrations as the red sequence begins to form. 

We explore the applicability of our methodology to colour agnostic cluster finders by performing the scaling relation measurement of VT clusters. The results are again consistent {with those obtained for the redMaPPer clusters in this redshift range}, as expected, indicating that our methods hold for {other} cluster selection algorithms. A clear result of our work is the recommendation that $\mu_{\star}$ be incorporated as the mass proxy for VT clusters.

\section*{Acknowledgments}

MESP has received partial support from the Conselho Nacional de Desenvolvimento Cient\'ifico e Tecnol\'ogico (CNPq), Brazil, and from the Fermilab Center for Particle Astrophysics. MESP thanks Dr Phil Marshall, Katalin Takats, Lucas Secco, Oleg Burgue\~no and Franco N. Bellomo for their help in developing portions of the codes during the \#AstroHackWeek unconference event.
MM is partially supported by CNPq (grant 312353/2015-4) and FAPERJ. 
Fora Temer.
AP acknowledges support from the URA research scholar award and the UCL PhD studentship. We thank Eli Rykoff for useful discussions. %

This work is based on observations obtained with MegaPrime/MegaCam, a joint project of CFHT and CEA/DAPNIA, at the Canada--France--Hawaii Telescope (CFHT), which is operated by the National Research Council (NRC) of Canada, the Institut National des Sciences de l'Univers of the Centre National de la Recherche Scientifique (CNRS) of France, and the University of Hawaii. The Brazilian partnership on CFHT is managed by the Laborat\'orio Nacional de Astrof\'isica (LNA). We thank the support of the Laborat\'orio Interinstitucional de e-Astronomia (LIneA). 
We thank the CFHTLenS team for their pipeline development and verification upon which much of the CS82 survey pipeline was built.

Funding for SDSS-III has been provided by the Alfred P. Sloan Foundation, the Participating Institutions, the National Science Foundation, and the U.S. Department of Energy Office of Science. The SDSS-III web site is http://www.sdss3.org/.

SDSS-III is managed by the Astrophysical Research Consortium for the Participating Institutions of the SDSS-III Collaboration including the University of Arizona, the Brazilian Participation Group, Brookhaven National Laboratory, Carnegie Mellon University, University of Florida, the French Participation Group, the German Participation Group, Harvard University, the Instituto de Astrofisica de Canarias, the Michigan State/Notre Dame/JINA Participation Group, Johns Hopkins University, Lawrence Berkeley National Laboratory, Max Planck Institute for Astrophysics, Max Planck Institute for Extraterrestrial Physics, New Mexico State University, New York University, Ohio State University, Pennsylvania State University, University of Portsmouth, Princeton University, the Spanish Participation Group, University of Tokyo, University of Utah, Vanderbilt University, University of Virginia, University of Washington, and Yale University. 

This manuscript has been authored by Fermi Research Alliance, LLC under Contract No. DE-AC02-07CH11359 with the U.S. Department of Energy, Office of Science, Office of High Energy Physics. The United States Government retains and the publisher, by accepting the article for publication, acknowledges that the United States Government retains a non-exclusive, paid-up, irrevocable, world-wide license to publish or reproduce the published form of this manuscript, or allow others to do so, for United States Government purposes.

\bibliographystyle{mnras}
\bibliography{vtproject} %

\begin{thebibliography}{}
\makeatletter
\relax
\def\mn@urlcharsother{\let\do\@makeother \do\$\do\&\do\#\do\^\do\_\do\%\do\~}
\def\mn@doi{\begingroup\mn@urlcharsother \@ifnextchar [ {\mn@doi@}
  {\mn@doi@[]}}
\def\mn@doi@[#1]#2{\def\@tempa{#1}\ifx\@tempa\@empty \href
  {http://dx.doi.org/#2} {doi:#2}\else \href {http://dx.doi.org/#2} {#1}\fi
  \endgroup}
\def\mn@eprint#1#2{\mn@eprint@#1:#2::\@nil}
\def\mn@eprint@arXiv#1{\href {http://arxiv.org/abs/#1} {{\tt arXiv:#1}}}
\def\mn@eprint@dblp#1{\href {http://dblp.uni-trier.de/rec/bibtex/#1.xml}
  {dblp:#1}}
\def\mn@eprint@#1:#2:#3:#4\@nil{\def\@tempa {#1}\def\@tempb {#2}\def\@tempc
  {#3}\ifx \@tempc \@empty \let \@tempc \@tempb \let \@tempb \@tempa \fi \ifx
  \@tempb \@empty \def\@tempb {arXiv}\fi \@ifundefined
  {mn@eprint@\@tempb}{\@tempb:\@tempc}{\expandafter \expandafter \csname
  mn@eprint@\@tempb\endcsname \expandafter{\@tempc}}}

\bibitem[\protect\citeauthoryear{{Aihara} et~al.,}{{Aihara}
  et~al.}{2011}]{2011ApJS..193...29A}
{Aihara} H.,  et~al., 2011, \mn@doi [\apjs] {10.1088/0067-0049/193/2/29}, \href
  {http://adsabs.harvard.edu/abs/2011ApJS..193...29A} {193, 29}

\bibitem[\protect\citeauthoryear{{Allen}, {Evrard}  \& {Mantz}}{{Allen}
  et~al.}{2011}]{2011ARA&A..49..409A}
{Allen} S.~W.,  {Evrard} A.~E.,   {Mantz} A.~B.,  2011, \mn@doi [\araa]
  {10.1146/annurev-astro-081710-102514}, \href
  {http://adsabs.harvard.edu/abs/2011ARA%26A..49..409A} {49, 409}

\bibitem[\protect\citeauthoryear{{Andreon}}{{Andreon}}{2012}]{2012A&A...548A..83A}
{Andreon} S.,  2012, \mn@doi [\aap] {10.1051/0004-6361/201220284}, \href
  {http://adsabs.harvard.edu/abs/2012A%26A...548A..83A} {548, A83}

\bibitem[\protect\citeauthoryear{{Annis} et~al.,}{{Annis}
  et~al.}{2014}]{2014ApJ...794..120A}
{Annis} J.,  et~al., 2014, \mn@doi [\apj] {10.1088/0004-637X/794/2/120}, \href
  {http://adsabs.harvard.edu/abs/2014ApJ...794..120A} {794, 120}

\bibitem[\protect\citeauthoryear{{Bartelmann}}{{Bartelmann}}{1996}]{1996A&A...313..697B}
{Bartelmann} M.,  1996, \aap, \href
  {http://adsabs.harvard.edu/abs/1996A%26A...313..697B} {313, 697}

\bibitem[\protect\citeauthoryear{{Battaglia} et~al.,}{{Battaglia}
  et~al.}{2016}]{2016JCAP...08..013B}
{Battaglia} N.,  et~al., 2016, \mn@doi [\jcap] {10.1088/1475-7516/2016/08/013},
  \href {http://adsabs.harvard.edu/abs/2016JCAP...08..013B} {8, 013}

\bibitem[\protect\citeauthoryear{{Ben{\'{\i}}tez}}{{Ben{\'{\i}}tez}}{2000}]{2000ApJ...536..571B}
{Ben{\'{\i}}tez} N.,  2000, \mn@doi [\apj] {10.1086/308947}, \href
  {http://adsabs.harvard.edu/abs/2000ApJ...536..571B} {536, 571}

\bibitem[\protect\citeauthoryear{{Benjamin} et~al.,}{{Benjamin}
  et~al.}{2013}]{2013MNRAS.431.1547B}
{Benjamin} J.,  et~al., 2013, \mn@doi [\mnras] {10.1093/mnras/stt276}, \href
  {http://adsabs.harvard.edu/abs/2013MNRAS.431.1547B} {431, 1547}

\bibitem[\protect\citeauthoryear{{Blanton} \& {Roweis}}{{Blanton} \&
  {Roweis}}{2007}]{2007AJ....133..734B}
{Blanton} M.~R.,  {Roweis} S.,  2007, \mn@doi [\aj] {10.1086/510127}, \href
  {http://adsabs.harvard.edu/abs/2007AJ....133..734B} {133, 734}

\bibitem[\protect\citeauthoryear{{Bundy} et~al.,}{{Bundy}
  et~al.}{2015}]{2015ApJS..221...15B}
{Bundy} K.,  et~al., 2015, \mn@doi [\apjs] {10.1088/0067-0049/221/1/15}, \href
  {http://adsabs.harvard.edu/abs/2015ApJS..221...15B} {221, 15}

\bibitem[\protect\citeauthoryear{{Clampitt} \& {Jain}}{{Clampitt} \&
  {Jain}}{2015}]{2015MNRAS.454.3357C}
{Clampitt} J.,  {Jain} B.,  2015, \mn@doi [\mnras] {10.1093/mnras/stv2215},
  \href {http://adsabs.harvard.edu/abs/2015MNRAS.454.3357C} {454, 3357}

\bibitem[\protect\citeauthoryear{{Coil} et~al.,}{{Coil} et~al.}{2011}]{PRIMUS}
{Coil} A.~L.,  et~al., 2011, \mn@doi [\apj] {10.1088/0004-637X/741/1/8}, \href
  {http://adsabs.harvard.edu/abs/2011ApJ...741....8C} {741, 8}

\bibitem[\protect\citeauthoryear{{Colless} et~al.,}{{Colless}
  et~al.}{2001}]{2dFGRS}
{Colless} M.,  et~al., 2001, \mn@doi [\mnras]
  {10.1046/j.1365-8711.2001.04902.x}, \href
  {http://adsabs.harvard.edu/abs/2001MNRAS.328.1039C} {328, 1039}

\bibitem[\protect\citeauthoryear{{Conroy}}{{Conroy}}{2013}]{2013ARA&A..51..393C}
{Conroy} C.,  2013, \mn@doi [\araa] {10.1146/annurev-astro-082812-141017},
  \href {http://adsabs.harvard.edu/abs/2013ARA%26A..51..393C} {51, 393}

\bibitem[\protect\citeauthoryear{{Conroy} \& {Gunn}}{{Conroy} \&
  {Gunn}}{2010}]{fsps}
{Conroy} C.,  {Gunn} J.~E.,  2010, \mn@doi [\apj]
  {10.1088/0004-637X/712/2/833}, \href
  {http://adsabs.harvard.edu/abs/2010ApJ...712..833C} {712, 833}

\bibitem[\protect\citeauthoryear{{Croom}, {Smith}, {Boyle}, {Shanks},
  {Loaring}, {Miller}  \& {Lewis}}{{Croom} et~al.}{2001}]{2QZ}
{Croom} S.~M.,  {Smith} R.~J.,  {Boyle} B.~J.,  {Shanks} T.,  {Loaring} N.~S.,
  {Miller} L.,   {Lewis} I.~J.,  2001, \mn@doi [\mnras]
  {10.1046/j.1365-8711.2001.04474.x}, \href
  {http://adsabs.harvard.edu/abs/2001MNRAS.322L..29C} {322, L29}

\bibitem[\protect\citeauthoryear{{Croom}, {Smith}, {Boyle}, {Shanks}, {Miller},
  {Outram}  \& {Loaring}}{{Croom} et~al.}{2004}]{2004MNRAS.349.1397C}
{Croom} S.~M.,  {Smith} R.~J.,  {Boyle} B.~J.,  {Shanks} T.,  {Miller} L.,
  {Outram} P.~J.,   {Loaring} N.~S.,  2004, \mn@doi [\mnras]
  {10.1111/j.1365-2966.2004.07619.x}, \href
  {http://adsabs.harvard.edu/abs/2004MNRAS.349.1397C} {349, 1397}

\bibitem[\protect\citeauthoryear{{Croom} et~al.,}{{Croom}
  et~al.}{2009a}]{2SLAQ}
{Croom} S.~M.,  et~al., 2009a, \mn@doi [\mnras]
  {10.1111/j.1365-2966.2008.14052.x}, \href
  {http://adsabs.harvard.edu/abs/2009MNRAS.392...19C} {392, 19}

\bibitem[\protect\citeauthoryear{{Croom} et~al.,}{{Croom}
  et~al.}{2009b}]{2009MNRAS.392...19C}
{Croom} S.~M.,  et~al., 2009b, \mn@doi [\mnras]
  {10.1111/j.1365-2966.2008.14052.x}, \href
  {http://adsabs.harvard.edu/abs/2009MNRAS.392...19C} {392, 19}

\bibitem[\protect\citeauthoryear{{Diemer}}{{Diemer}}{2015}]{2015ascl.soft01016D}
{Diemer} B.,  2015, {Colossus: COsmology, haLO, and large-Scale StrUcture
  toolS}, Astrophysics Source Code Library (\mn@eprint {ascl} {1501.016})

\bibitem[\protect\citeauthoryear{{Drinkwater} et~al.,}{{Drinkwater}
  et~al.}{2010}]{wigglez}
{Drinkwater} M.~J.,  et~al., 2010, \mn@doi [\mnras]
  {10.1111/j.1365-2966.2009.15754.x}, \href
  {http://adsabs.harvard.edu/abs/2010MNRAS.401.1429D} {401, 1429}

\bibitem[\protect\citeauthoryear{{Duffy}, {Schaye}, {Kay}  \& {Dalla
  Vecchia}}{{Duffy} et~al.}{2008}]{2008MNRAS.390L..64D}
{Duffy} A.~R.,  {Schaye} J.,  {Kay} S.~T.,   {Dalla Vecchia} C.,  2008, \mn@doi
  [\mnras] {10.1111/j.1745-3933.2008.00537.x}, \href
  {http://adsabs.harvard.edu/abs/2008MNRAS.390L..64D} {390, L64}

\bibitem[\protect\citeauthoryear{{Eisenstein} et~al.,}{{Eisenstein}
  et~al.}{2011}]{2011AJ....142...72E}
{Eisenstein} D.~J.,  et~al., 2011, \mn@doi [\aj] {10.1088/0004-6256/142/3/72},
  \href {http://adsabs.harvard.edu/abs/2011AJ....142...72E} {142, 72}

\bibitem[\protect\citeauthoryear{{Erben} et~al.,}{{Erben}
  et~al.}{2013}]{2013MNRAS.433.2545E}
{Erben} T.,  et~al., 2013, \mn@doi [\mnras] {10.1093/mnras/stt928}, \href
  {http://adsabs.harvard.edu/abs/2013MNRAS.433.2545E} {433, 2545}

\bibitem[\protect\citeauthoryear{{Erben} et~al.}{{Erben}
  et~al.}{2017}]{erben2017}
{Erben} T.,  et~al., 2017, In preparation

\bibitem[\protect\citeauthoryear{{Ettori} \& {Meneghetti}}{{Ettori} \&
  {Meneghetti}}{2013}]{2013SSRv..177....1E}
{Ettori} S.,  {Meneghetti} M.,  2013, \mn@doi [\ssr]
  {10.1007/s11214-013-0002-x}, \href
  {http://adsabs.harvard.edu/abs/2013SSRv..177....1E} {177, 1}

\bibitem[\protect\citeauthoryear{{Ford} \& {VanderPlas}}{{Ford} \&
  {VanderPlas}}{2016}]{2016AJ....152..228F}
{Ford} J.,  {VanderPlas} J.,  2016, \mn@doi [\aj]
  {10.3847/1538-3881/152/6/228}, \href
  {http://adsabs.harvard.edu/abs/2016AJ....152..228F} {152, 228}

\bibitem[\protect\citeauthoryear{{Ford}, {Hildebrandt}, {Van Waerbeke},
  {Erben}, {Laigle}, {Milkeraitis}  \& {Morrison}}{{Ford}
  et~al.}{2014}]{2014MNRAS.439.3755F}
{Ford} J.,  {Hildebrandt} H.,  {Van Waerbeke} L.,  {Erben} T.,  {Laigle} C.,
  {Milkeraitis} M.,   {Morrison} C.~B.,  2014, \mn@doi [\mnras]
  {10.1093/mnras/stu225}, \href
  {http://adsabs.harvard.edu/abs/2014MNRAS.439.3755F} {439, 3755}

\bibitem[\protect\citeauthoryear{{Ford} et~al.,}{{Ford}
  et~al.}{2015}]{2015MNRAS.447.1304F}
{Ford} J.,  et~al., 2015, \mn@doi [\mnras] {10.1093/mnras/stu2545}, \href
  {http://adsabs.harvard.edu/abs/2015MNRAS.447.1304F} {447, 1304}

\bibitem[\protect\citeauthoryear{{Frieman} et~al.,}{{Frieman}
  et~al.}{2008}]{2008AJ....135..338F}
{Frieman} J.~A.,  et~al., 2008, \mn@doi [\aj] {10.1088/0004-6256/135/1/338},
  \href {http://adsabs.harvard.edu/abs/2008AJ....135..338F} {135, 338}

\bibitem[\protect\citeauthoryear{{Garilli} et~al.,}{{Garilli}
  et~al.}{2008}]{Garilli08}
{Garilli} B.,  et~al., 2008, \mn@doi [\aap] {10.1051/0004-6361:20078878}, \href
  {http://adsabs.harvard.edu/abs/2008A%26A...486..683G} {486, 683}

\bibitem[\protect\citeauthoryear{{Geach} et~al.,}{{Geach}
  et~al.}{2017}]{2017arXiv170505451G}
{Geach} J.~E.,  et~al., 2017, preprint, \href
  {http://adsabs.harvard.edu/abs/2017arXiv170505451G} {} (\mn@eprint {arXiv}
  {1705.05451})

\bibitem[\protect\citeauthoryear{{Girardi}, {Bressan}, {Bertelli}  \&
  {Chiosi}}{{Girardi} et~al.}{2000}]{padova1}
{Girardi} L.,  {Bressan} A.,  {Bertelli} G.,   {Chiosi} C.,  2000, \mn@doi
  [\aaps] {10.1051/aas:2000126}, \href
  {http://adsabs.harvard.edu/abs/2000A%26AS..141..371G} {141, 371}

\bibitem[\protect\citeauthoryear{{Gonzalez}, {Rodriguez}, {Garc{\'{\i}}a
  Lambas}, {Merch{\'a}n}, {Fo{\"e}x}  \& {Chalela}}{{Gonzalez}
  et~al.}{2017}]{2017MNRAS.465.1348G}
{Gonzalez} E.~J.,  {Rodriguez} F.,  {Garc{\'{\i}}a Lambas} D.,  {Merch{\'a}n}
  M.,  {Fo{\"e}x} G.,   {Chalela} M.,  2017, \mn@doi [\mnras]
  {10.1093/mnras/stw2803}, \href
  {http://adsabs.harvard.edu/abs/2017MNRAS.465.1348G} {465, 1348}

\bibitem[\protect\citeauthoryear{{Haiman}, {Mohr}  \& {Holder}}{{Haiman}
  et~al.}{2001}]{2001ApJ...553..545H}
{Haiman} Z.,  {Mohr} J.~J.,   {Holder} G.~P.,  2001, \mn@doi [\apj]
  {10.1086/320939}, \href {http://adsabs.harvard.edu/abs/2001ApJ...553..545H}
  {553, 545}

\bibitem[\protect\citeauthoryear{{Hand} et~al.,}{{Hand}
  et~al.}{2015}]{2015PhRvD..91f2001H}
{Hand} N.,  et~al., 2015, \mn@doi [\prd] {10.1103/PhysRevD.91.062001}, \href
  {http://adsabs.harvard.edu/abs/2015PhRvD..91f2001H} {91, 062001}

\bibitem[\protect\citeauthoryear{{Hao} et~al.,}{{Hao}
  et~al.}{2010}]{2010ApJS..191..254H}
{Hao} J.,  et~al., 2010, \mn@doi [\apjs] {10.1088/0067-0049/191/2/254}, \href
  {http://adsabs.harvard.edu/abs/2010ApJS..191..254H} {191, 254}

\bibitem[\protect\citeauthoryear{{Harvey}, {Massey}, {Kitching}, {Taylor}  \&
  {Tittley}}{{Harvey} et~al.}{2015}]{2015Sci...347.1462H}
{Harvey} D.,  {Massey} R.,  {Kitching} T.,  {Taylor} A.,   {Tittley} E.,  2015,
  \mn@doi [Science] {10.1126/science.1261381}, \href
  {http://adsabs.harvard.edu/abs/2015Sci...347.1462H} {347, 1462}

\bibitem[\protect\citeauthoryear{{Heymans} et~al.,}{{Heymans}
  et~al.}{2012}]{2012MNRAS.427..146H}
{Heymans} C.,  et~al., 2012, \mn@doi [\mnras]
  {10.1111/j.1365-2966.2012.21952.x}, \href
  {http://adsabs.harvard.edu/abs/2012MNRAS.427..146H} {427, 146}

\bibitem[\protect\citeauthoryear{{Hildebrandt} et~al.,}{{Hildebrandt}
  et~al.}{2012}]{2012MNRAS.421.2355H}
{Hildebrandt} H.,  et~al., 2012, \mn@doi [\mnras]
  {10.1111/j.1365-2966.2012.20468.x}, \href
  {http://adsabs.harvard.edu/abs/2012MNRAS.421.2355H} {421, 2355}

\bibitem[\protect\citeauthoryear{{Hoeting}, {Madigan}, {Raftery}  \&
  {Volinsky}}{{Hoeting} et~al.}{1999}]{hoeting}
{Hoeting} J.~A.,  {Madigan} D.,  {Raftery} A.~E.,   {Volinsky} C.~T.,  1999,
  Statist. Sci., 14, 382

\bibitem[\protect\citeauthoryear{{Hudson} et~al.,}{{Hudson}
  et~al.}{2015}]{2015MNRAS.447..298H}
{Hudson} M.~J.,  et~al., 2015, \mn@doi [\mnras] {10.1093/mnras/stu2367}, \href
  {http://adsabs.harvard.edu/abs/2015MNRAS.447..298H} {447, 298}

\bibitem[\protect\citeauthoryear{{Jarvis} et~al.,}{{Jarvis}
  et~al.}{2016}]{2016MNRAS.460.2245J}
{Jarvis} M.,  et~al., 2016, \mn@doi [\mnras] {10.1093/mnras/stw990}, \href
  {http://adsabs.harvard.edu/abs/2016MNRAS.460.2245J} {460, 2245}

\bibitem[\protect\citeauthoryear{{Johnston} et~al.,}{{Johnston}
  et~al.}{2007}]{2007arXiv0709.1159J}
{Johnston} D.~E.,  et~al., 2007, preprint, \href
  {http://adsabs.harvard.edu/abs/2007arXiv0709.1159J} {} (\mn@eprint {arXiv}
  {0709.1159})

\bibitem[\protect\citeauthoryear{{Jones} et~al.,}{{Jones} et~al.}{2009}]{6dF}
{Jones} D.~H.,  et~al., 2009, \mn@doi [\mnras]
  {10.1111/j.1365-2966.2009.15338.x}, \href
  {http://adsabs.harvard.edu/abs/2009MNRAS.399..683J} {399, 683}

\bibitem[\protect\citeauthoryear{{Kettula} et~al.,}{{Kettula}
  et~al.}{2015}]{2015MNRAS.451.1460K}
{Kettula} K.,  et~al., 2015, \mn@doi [\mnras] {10.1093/mnras/stv923}, \href
  {http://adsabs.harvard.edu/abs/2015MNRAS.451.1460K} {451, 1460}

\bibitem[\protect\citeauthoryear{{Kitching}, {Miller}, {Heymans}, {van
  Waerbeke}  \& {Heavens}}{{Kitching} et~al.}{2008}]{2008MNRAS.390..149K}
{Kitching} T.~D.,  {Miller} L.,  {Heymans} C.~E.,  {van Waerbeke} L.,
  {Heavens} A.~F.,  2008, \mn@doi [\mnras] {10.1111/j.1365-2966.2008.13628.x},
  \href {http://adsabs.harvard.edu/abs/2008MNRAS.390..149K} {390, 149}

\bibitem[\protect\citeauthoryear{{Kitching} et~al.,}{{Kitching}
  et~al.}{2012}]{2012MNRAS.423.3163K}
{Kitching} T.~D.,  et~al., 2012, \mn@doi [\mnras]
  {10.1111/j.1365-2966.2012.21095.x}, \href
  {http://adsabs.harvard.edu/abs/2012MNRAS.423.3163K} {423, 3163}

\bibitem[\protect\citeauthoryear{{Koester} et~al.,}{{Koester}
  et~al.}{2007}]{2007ApJ...660..239K}
{Koester} B.~P.,  et~al., 2007, \mn@doi [\apj] {10.1086/509599}, \href
  {http://adsabs.harvard.edu/abs/2007ApJ...660..239K} {660, 239}

\bibitem[\protect\citeauthoryear{{Kravtsov} \& {Borgani}}{{Kravtsov} \&
  {Borgani}}{2012}]{2012ARA&A..50..353K}
{Kravtsov} A.~V.,  {Borgani} S.,  2012, \mn@doi [\araa]
  {10.1146/annurev-astro-081811-125502}, \href
  {http://adsabs.harvard.edu/abs/2012ARA%26A..50..353K} {50, 353}

\bibitem[\protect\citeauthoryear{{Kuijken} et~al.,}{{Kuijken}
  et~al.}{2015}]{2015MNRAS.454.3500K}
{Kuijken} K.,  et~al., 2015, \mn@doi [\mnras] {10.1093/mnras/stv2140}, \href
  {http://adsabs.harvard.edu/abs/2015MNRAS.454.3500K} {454, 3500}

\bibitem[\protect\citeauthoryear{{LaMassa} et~al.,}{{LaMassa}
  et~al.}{2016}]{S82X21}
{LaMassa} S.~M.,  et~al., 2016, \mn@doi [\apj] {10.3847/0004-637X/817/2/172},
  \href {http://adsabs.harvard.edu/abs/2016ApJ...817..172L} {817, 172}

\bibitem[\protect\citeauthoryear{{Lawrence} et~al.,}{{Lawrence}
  et~al.}{2007}]{UKIDSS}
{Lawrence} A.,  et~al., 2007, \mn@doi [\mnras]
  {10.1111/j.1365-2966.2007.12040.x}, \href
  {http://adsabs.harvard.edu/abs/2007MNRAS.379.1599L} {379, 1599}

\bibitem[\protect\citeauthoryear{{Le F{\`e}vre} et~al.,}{{Le F{\`e}vre}
  et~al.}{2013}]{2013A&A...559A..14L}
{Le F{\`e}vre} O.,  et~al., 2013, \mn@doi [\aap] {10.1051/0004-6361/201322179},
  \href {http://adsabs.harvard.edu/abs/2013A%26A...559A..14L} {559, A14}

\bibitem[\protect\citeauthoryear{{Leauthaud} et~al.,}{{Leauthaud}
  et~al.}{2017}]{2017MNRAS.467.3024L}
{Leauthaud} A.,  et~al., 2017, \mn@doi [\mnras] {10.1093/mnras/stx258}, \href
  {http://adsabs.harvard.edu/abs/2017MNRAS.467.3024L} {467, 3024}

\bibitem[\protect\citeauthoryear{{Li} et~al.,}{{Li}
  et~al.}{2014}]{2014MNRAS.438.2864L}
{Li} R.,  et~al., 2014, \mn@doi [\mnras] {10.1093/mnras/stt2395}, \href
  {http://adsabs.harvard.edu/abs/2014MNRAS.438.2864L} {438, 2864}

\bibitem[\protect\citeauthoryear{{Li} et~al.,}{{Li}
  et~al.}{2016}]{2016MNRAS.458.2573L}
{Li} R.,  et~al., 2016, \mn@doi [\mnras] {10.1093/mnras/stw494}, \href
  {http://adsabs.harvard.edu/abs/2016MNRAS.458.2573L} {458, 2573}

\bibitem[\protect\citeauthoryear{{Liu} et~al.,}{{Liu}
  et~al.}{2015}]{2015MNRAS.450.2888L}
{Liu} X.,  et~al., 2015, \mn@doi [\mnras] {10.1093/mnras/stv784}, \href
  {http://adsabs.harvard.edu/abs/2015MNRAS.450.2888L} {450, 2888}

\bibitem[\protect\citeauthoryear{{Mandelbaum}, {Seljak}, {Baldauf}  \&
  {Smith}}{{Mandelbaum} et~al.}{2010}]{2010MNRAS.405.2078M}
{Mandelbaum} R.,  {Seljak} U.,  {Baldauf} T.,   {Smith} R.~E.,  2010, \mn@doi
  [\mnras] {10.1111/j.1365-2966.2010.16619.x}, \href
  {http://adsabs.harvard.edu/abs/2010MNRAS.405.2078M} {405, 2078}

\bibitem[\protect\citeauthoryear{{Marigo} \& {Girardi}}{{Marigo} \&
  {Girardi}}{2007}]{padova2}
{Marigo} P.,  {Girardi} L.,  2007, \mn@doi [\aap] {10.1051/0004-6361:20066772},
  \href {http://adsabs.harvard.edu/abs/2007A%26A...469..239M} {469, 239}

\bibitem[\protect\citeauthoryear{{Marigo}, {Girardi}, {Bressan}, {Groenewegen},
  {Silva}  \& {Granato}}{{Marigo} et~al.}{2008}]{padova3}
{Marigo} P.,  {Girardi} L.,  {Bressan} A.,  {Groenewegen} M.~A.~T.,  {Silva}
  L.,   {Granato} G.~L.,  2008, \mn@doi [\aap] {10.1051/0004-6361:20078467},
  \href {http://adsabs.harvard.edu/abs/2008A%26A...482..883M} {482, 883}

\bibitem[\protect\citeauthoryear{{Melchior}, {Sutter}, {Sheldon}, {Krause}  \&
  {Wandelt}}{{Melchior} et~al.}{2014}]{2014MNRAS.440.2922M}
{Melchior} P.,  {Sutter} P.~M.,  {Sheldon} E.~S.,  {Krause} E.,   {Wandelt}
  B.~D.,  2014, \mn@doi [\mnras] {10.1093/mnras/stu456}, \href
  {http://adsabs.harvard.edu/abs/2014MNRAS.440.2922M} {440, 2922}

\bibitem[\protect\citeauthoryear{{Melchior} et~al.,}{{Melchior}
  et~al.}{2017}]{2016arXiv161006890M}
{Melchior} P.,  et~al., 2017, \mn@doi [\mnras] {10.1093/mnras/stx1053}, \href
  {http://adsabs.harvard.edu/abs/2017MNRAS.469.4899M} {469, 4899}

\bibitem[\protect\citeauthoryear{{Menci}, {Grazian}, {Castellano}  \&
  {Sanchez}}{{Menci} et~al.}{2016}]{2016ApJ...825L...1M}
{Menci} N.,  {Grazian} A.,  {Castellano} M.,   {Sanchez} N.~G.,  2016, \mn@doi
  [\apjl] {10.3847/2041-8205/825/1/L1}, \href
  {http://adsabs.harvard.edu/abs/2016ApJ...825L...1M} {825, L1}

\bibitem[\protect\citeauthoryear{{Miller}, {Kitching}, {Heymans}, {Heavens}  \&
  {van Waerbeke}}{{Miller} et~al.}{2007}]{2007MNRAS.382..315M}
{Miller} L.,  {Kitching} T.~D.,  {Heymans} C.,  {Heavens} A.~F.,   {van
  Waerbeke} L.,  2007, \mn@doi [\mnras] {10.1111/j.1365-2966.2007.12363.x},
  \href {http://adsabs.harvard.edu/abs/2007MNRAS.382..315M} {382, 315}

\bibitem[\protect\citeauthoryear{{Miller} et~al.,}{{Miller}
  et~al.}{2013}]{2013MNRAS.429.2858M}
{Miller} L.,  et~al., 2013, \mn@doi [\mnras] {10.1093/mnras/sts454}, \href
  {http://adsabs.harvard.edu/abs/2013MNRAS.429.2858M} {429, 2858}

\bibitem[\protect\citeauthoryear{{Miralda-Escude}}{{Miralda-Escude}}{1991}]{1991ApJ...370....1M}
{Miralda-Escude} J.,  1991, \mn@doi [\apj] {10.1086/169789}, \href
  {http://adsabs.harvard.edu/abs/1991ApJ...370....1M} {370, 1}

\bibitem[\protect\citeauthoryear{{Moraes} et~al.,}{{Moraes}
  et~al.}{2014}]{2014RMxAC..44..202M}
{Moraes} B.,  et~al., 2014, in Revista Mexicana de Astronomia y Astrofisica
  Conference Series. pp 202--203

\bibitem[\protect\citeauthoryear{{Navarro}, {Frenk}  \& {White}}{{Navarro}
  et~al.}{1996}]{1996ApJ...462..563N}
{Navarro} J.~F.,  {Frenk} C.~S.,   {White} S.~D.~M.,  1996, \mn@doi [\apj]
  {10.1086/177173}, \href {http://adsabs.harvard.edu/abs/1996ApJ...462..563N}
  {462, 563}

\bibitem[\protect\citeauthoryear{{Newman} et~al.,}{{Newman}
  et~al.}{2012}]{DEEP2}
{Newman} J.~A.,  et~al., 2012, preprint, \href
  {http://adsabs.harvard.edu/abs/2012arXiv1203.3192N} {} (\mn@eprint {arXiv}
  {1203.3192})

\bibitem[\protect\citeauthoryear{{Niemiec} et~al.,}{{Niemiec}
  et~al.}{2017}]{2017arXiv170303348N}
{Niemiec} A.,  et~al., 2017, preprint, \href
  {http://adsabs.harvard.edu/abs/2017arXiv170303348N} {} (\mn@eprint {arXiv}
  {1703.03348})

\bibitem[\protect\citeauthoryear{{Oguri}}{{Oguri}}{2014}]{2014MNRAS.444..147O}
{Oguri} M.,  2014, \mn@doi [\mnras] {10.1093/mnras/stu1446}, \href
  {http://adsabs.harvard.edu/abs/2014MNRAS.444..147O} {444, 147}

\bibitem[\protect\citeauthoryear{{Palmese} \& {DES Collaboration}}{{Palmese} \&
  {DES Collaboration}}{2017}]{palmese2017}
{Palmese} A.,  {DES Collaboration} 2017, In preparation

\bibitem[\protect\citeauthoryear{{Palmese} et~al.,}{{Palmese}
  et~al.}{2016}]{2016MNRAS.463.1486P}
{Palmese} A.,  et~al., 2016, \mn@doi [\mnras] {10.1093/mnras/stw2062}, \href
  {http://adsabs.harvard.edu/abs/2016MNRAS.463.1486P} {463, 1486}

\bibitem[\protect\citeauthoryear{{Penna-Lima}, {Makler}  \&
  {Wuensche}}{{Penna-Lima} et~al.}{2014}]{2014JCAP...05..039P}
{Penna-Lima} M.,  {Makler} M.,   {Wuensche} C.~A.,  2014, \mn@doi [\jcap]
  {10.1088/1475-7516/2014/05/039}, \href
  {http://adsabs.harvard.edu/abs/2014JCAP...05..039P} {5, 039}

\bibitem[\protect\citeauthoryear{{Pizzuti} et~al.,}{{Pizzuti}
  et~al.}{2016}]{2016JCAP...04..023P}
{Pizzuti} L.,  et~al., 2016, \mn@doi [\jcap] {10.1088/1475-7516/2016/04/023},
  \href {http://adsabs.harvard.edu/abs/2016JCAP...04..023P} {4, 023}

\bibitem[\protect\citeauthoryear{{Reis} et~al.,}{{Reis}
  et~al.}{2012}]{2012ApJ...747...59R}
{Reis} R.~R.~R.,  et~al., 2012, \mn@doi [\apj] {10.1088/0004-637X/747/1/59},
  \href {http://adsabs.harvard.edu/abs/2012ApJ...747...59R} {747, 59}

\bibitem[\protect\citeauthoryear{{Reyes}, {Mandelbaum}, {Gunn}, {Nakajima},
  {Seljak}  \& {Hirata}}{{Reyes} et~al.}{2012}]{2012MNRAS.425.2610R}
{Reyes} R.,  {Mandelbaum} R.,  {Gunn} J.~E.,  {Nakajima} R.,  {Seljak} U.,
  {Hirata} C.~M.,  2012, \mn@doi [\mnras] {10.1111/j.1365-2966.2012.21472.x},
  \href {http://adsabs.harvard.edu/abs/2012MNRAS.425.2610R} {425, 2610}

\bibitem[\protect\citeauthoryear{{Roediger} et~al.,}{{Roediger}
  et~al.}{2017}]{2017ApJ...836..120R}
{Roediger} J.~C.,  et~al., 2017, \mn@doi [\apj] {10.3847/1538-4357/836/1/120},
  \href {http://adsabs.harvard.edu/abs/2017ApJ...836..120R} {836, 120}

\bibitem[\protect\citeauthoryear{{Rozo} \& {Rykoff}}{{Rozo} \&
  {Rykoff}}{2014}]{2014ApJ...783...80R}
{Rozo} E.,  {Rykoff} E.~S.,  2014, \mn@doi [\apj] {10.1088/0004-637X/783/2/80},
  \href {http://adsabs.harvard.edu/abs/2014ApJ...783...80R} {783, 80}

\bibitem[\protect\citeauthoryear{{Rykoff} et~al.,}{{Rykoff}
  et~al.}{2012}]{2012ApJ...746..178R}
{Rykoff} E.~S.,  et~al., 2012, \mn@doi [\apj] {10.1088/0004-637X/746/2/178},
  \href {http://adsabs.harvard.edu/abs/2012ApJ...746..178R} {746, 178}

\bibitem[\protect\citeauthoryear{{Rykoff} et~al.,}{{Rykoff}
  et~al.}{2014}]{2014ApJ...785..104R}
{Rykoff} E.~S.,  et~al., 2014, \mn@doi [\apj] {10.1088/0004-637X/785/2/104},
  \href {http://adsabs.harvard.edu/abs/2014ApJ...785..104R} {785, 104}

\bibitem[\protect\citeauthoryear{{Rykoff} et~al.,}{{Rykoff}
  et~al.}{2016}]{2016ApJS..224....1R}
{Rykoff} E.~S.,  et~al., 2016, \mn@doi [\apjs] {10.3847/0067-0049/224/1/1},
  \href {http://adsabs.harvard.edu/abs/2016ApJS..224....1R} {224, 1}

\bibitem[\protect\citeauthoryear{{S{\'a}nchez-Bl{\'a}zquez}
  et~al.,}{{S{\'a}nchez-Bl{\'a}zquez} et~al.}{2006}]{miles}
{S{\'a}nchez-Bl{\'a}zquez} P.,  et~al., 2006, \mn@doi [\mnras]
  {10.1111/j.1365-2966.2006.10699.x}, \href
  {http://adsabs.harvard.edu/abs/2006MNRAS.371..703S} {371, 703}

\bibitem[\protect\citeauthoryear{{Saro} et~al.,}{{Saro}
  et~al.}{2015}]{2015MNRAS.454.2305S}
{Saro} A.,  et~al., 2015, \mn@doi [\mnras] {10.1093/mnras/stv2141}, \href
  {http://adsabs.harvard.edu/abs/2015MNRAS.454.2305S} {454, 2305}

\bibitem[\protect\citeauthoryear{{Schneider}}{{Schneider}}{2005}]{2005astro.ph..9252S}
{Schneider} P.,  2005, ArXiv Astrophysics e-prints, \href
  {http://adsabs.harvard.edu/abs/2005astro.ph..9252S} {}

\bibitem[\protect\citeauthoryear{{Shan} et~al.,}{{Shan}
  et~al.}{2014}]{2014MNRAS.442.2534S}
{Shan} H.~Y.,  et~al., 2014, \mn@doi [\mnras] {10.1093/mnras/stu1040}, \href
  {http://adsabs.harvard.edu/abs/2014MNRAS.442.2534S} {442, 2534}

\bibitem[\protect\citeauthoryear{{Shan} et~al.,}{{Shan}
  et~al.}{2017}]{2017ApJ...840..104S}
{Shan} H.,  et~al., 2017, \mn@doi [\apj] {10.3847/1538-4357/aa6c68}, \href
  {http://adsabs.harvard.edu/abs/2017ApJ...840..104S} {840, 104}

\bibitem[\protect\citeauthoryear{{Sheldon} et~al.,}{{Sheldon}
  et~al.}{2001}]{2001ApJ...554..881S}
{Sheldon} E.~S.,  et~al., 2001, \mn@doi [\apj] {10.1086/321395}, \href
  {http://adsabs.harvard.edu/abs/2001ApJ...554..881S} {554, 881}

\bibitem[\protect\citeauthoryear{{Simet} et~al.,}{{Simet}
  et~al.}{2012}]{2012ApJ...748..128S}
{Simet} M.,  et~al., 2012, \mn@doi [\apj] {10.1088/0004-637X/748/2/128}, \href
  {http://adsabs.harvard.edu/abs/2012ApJ...748..128S} {748, 128}

\bibitem[\protect\citeauthoryear{{Simet}, {McClintock}, {Mandelbaum}, {Rozo},
  {Rykoff}, {Sheldon}  \& {Wechsler}}{{Simet}
  et~al.}{2017}]{2017MNRAS.466.3103S}
{Simet} M.,  {McClintock} T.,  {Mandelbaum} R.,  {Rozo} E.,  {Rykoff} E.,
  {Sheldon} E.,   {Wechsler} R.~H.,  2017, \mn@doi [\mnras]
  {10.1093/mnras/stw3250}, \href
  {http://adsabs.harvard.edu/abs/2017MNRAS.466.3103S} {466, 3103}

\bibitem[\protect\citeauthoryear{{Simha}, {Weinberg}, {Conroy}, {Dave},
  {Fardal}, {Katz}  \& {Oppenheimer}}{{Simha} et~al.}{2014}]{simha}
{Simha} V.,  {Weinberg} D.~H.,  {Conroy} C.,  {Dave} R.,  {Fardal} M.,  {Katz}
  N.,   {Oppenheimer} B.~D.,  2014, preprint, \href
  {http://adsabs.harvard.edu/abs/2014arXiv1404.0402S} {} (\mn@eprint {arXiv}
  {1404.0402})

\bibitem[\protect\citeauthoryear{{Soares-Santos} et~al.,}{{Soares-Santos}
  et~al.}{2011}]{2011ApJ...727...45S}
{Soares-Santos} M.,  et~al., 2011, \mn@doi [\apj] {10.1088/0004-637X/727/1/45},
  \href {http://adsabs.harvard.edu/abs/2011ApJ...727...45S} {727, 45}

\bibitem[\protect\citeauthoryear{{Soo} et~al.,}{{Soo} et~al.}{2017}]{Soo17}
{Soo} J.~Y.~H.,  et~al., 2017, preprint, \href
  {http://adsabs.harvard.edu/abs/2017arXiv170703169S} {} (\mn@eprint {arXiv}
  {1707.03169})

\bibitem[\protect\citeauthoryear{{Timlin} et~al.,}{{Timlin}
  et~al.}{2016}]{SpIES1}
{Timlin} J.~D.,  et~al., 2016, \mn@doi [\apjs] {10.3847/0067-0049/225/1/1},
  \href {http://adsabs.harvard.edu/abs/2016ApJS..225....1T} {225, 1}

\bibitem[\protect\citeauthoryear{{Velander} et~al.,}{{Velander}
  et~al.}{2014}]{2014MNRAS.437.2111V}
{Velander} M.,  et~al., 2014, \mn@doi [\mnras] {10.1093/mnras/stt2013}, \href
  {http://adsabs.harvard.edu/abs/2014MNRAS.437.2111V} {437, 2111}

\bibitem[\protect\citeauthoryear{{Vitorelli} et~al.}{{Vitorelli}
  et~al.}{2017}]{vitorelli2017}
{Vitorelli} A.~Z.,  et~al., 2017, In preparation

\bibitem[\protect\citeauthoryear{Voit}{Voit}{2005}]{RevModPhys.77.207}
Voit G.~M.,  2005, \mn@doi [Rev. Mod. Phys.] {10.1103/RevModPhys.77.207}, 77,
  207

\bibitem[\protect\citeauthoryear{{Welch} \& {DES Collaboration}}{{Welch} \&
  {DES Collaboration}}{2017}]{welch2017}
{Welch} B.,  {DES Collaboration} 2017, In preparation

\bibitem[\protect\citeauthoryear{{Wen} \& {Han}}{{Wen} \&
  {Han}}{2015}]{2015ApJ...807..178W}
{Wen} Z.~L.,  {Han} J.~L.,  2015, \mn@doi [\apj] {10.1088/0004-637X/807/2/178},
  \href {http://adsabs.harvard.edu/abs/2015ApJ...807..178W} {807, 178}

\bibitem[\protect\citeauthoryear{{Wiesner}, {Lin}  \&
  {Soares-Santos}}{{Wiesner} et~al.}{2015}]{2015MNRAS.452..701W}
{Wiesner} M.~P.,  {Lin} H.,   {Soares-Santos} M.,  2015, \mn@doi [\mnras]
  {10.1093/mnras/stv1332}, \href
  {http://adsabs.harvard.edu/abs/2015MNRAS.452..701W} {452, 701}

\bibitem[\protect\citeauthoryear{{Wright} \& {Brainerd}}{{Wright} \&
  {Brainerd}}{1999}]{1999astro.ph..8213O}
{Wright} C.~O.,  {Brainerd} T.~G.,  1999, ArXiv Astrophysics e-prints, \href
  {http://adsabs.harvard.edu/abs/1999astro.ph..8213O} {}

\bibitem[\protect\citeauthoryear{{Yang}, {Mo}, {van den Bosch}, {Jing},
  {Weinmann}  \& {Meneghetti}}{{Yang} et~al.}{2006}]{2006MNRAS.373.1159Y}
{Yang} X.,  {Mo} H.~J.,  {van den Bosch} F.~C.,  {Jing} Y.~P.,  {Weinmann}
  S.~M.,   {Meneghetti} M.,  2006, \mn@doi [\mnras]
  {10.1111/j.1365-2966.2006.11091.x}, \href
  {http://adsabs.harvard.edu/abs/2006MNRAS.373.1159Y} {373, 1159}

\bibitem[\protect\citeauthoryear{{Zitrin}, {Bartelmann}, {Umetsu}, {Oguri}  \&
  {Broadhurst}}{{Zitrin} et~al.}{2012}]{2012MNRAS.426.2944Z}
{Zitrin} A.,  {Bartelmann} M.,  {Umetsu} K.,  {Oguri} M.,   {Broadhurst} T.,
  2012, \mn@doi [\mnras] {10.1111/j.1365-2966.2012.21886.x}, \href
  {http://adsabs.harvard.edu/abs/2012MNRAS.426.2944Z} {426, 2944}

\bibitem[\protect\citeauthoryear{{de Jong} et~al.,}{{de Jong}
  et~al.}{2013}]{2013Msngr.154...44J}
{de Jong} J.~T.~A.,  et~al., 2013, The Messenger, \href
  {http://adsabs.harvard.edu/abs/2013Msngr.154...44J} {154, 44}

\bibitem[\protect\citeauthoryear{{de la Torre} et~al.,}{{de la Torre}
  et~al.}{2013}]{VIPERS}
{de la Torre} S.,  et~al., 2013, \mn@doi [\aap] {10.1051/0004-6361/201321463},
  \href {http://adsabs.harvard.edu/abs/2013A%26A...557A..54D} {557, A54}

\bibitem[\protect\citeauthoryear{{van Uitert}, {Hoekstra}, {Schrabback},
  {Gilbank}, {Gladders}  \& {Yee}}{{van Uitert}
  et~al.}{2012}]{2012A&A...545A..71V}
{van Uitert} E.,  {Hoekstra} H.,  {Schrabback} T.,  {Gilbank} D.~G.,
  {Gladders} M.~D.,   {Yee} H.~K.~C.,  2012, \mn@doi [\aap]
  {10.1051/0004-6361/201219295}, \href
  {http://adsabs.harvard.edu/abs/2012A%26A...545A..71V} {545, A71}

\makeatother
\end{thebibliography}

\bsp	%
\label{lastpage}
\end{document}